\documentclass[reprint,aps,nofootinbib,superscriptaddress]{revtex4-2}
\usepackage[utf8]{inputenc}
\usepackage[T1]{fontenc}
\usepackage{amsmath,amssymb,graphicx, epstopdf,enumerate,amsfonts,booktabs,color,soul}
\usepackage{slashed}
\usepackage{float,placeins}
\usepackage{appendix}
\usepackage{dcolumn}
\usepackage{relsize}
\usepackage{xcolor}
\usepackage[caption=false]{subfig}
\usepackage{hyperref}
\usepackage{xspace}
\usepackage{bm}
\usepackage{comment} 
\begin{document}

\title{Page Curve of Effective Hawking Radiation}

\author{Chia-Jui Chou}
\email{chiajuichou@nycu.edu.tw}
\affiliation{Department of Electrophysics, National Yang Ming Chiao Tung University, Hsinchu, ROC}

\author{Hans B. Lao}
\email{hanslao.sc07@nycu.edu.tw}
\affiliation{Department of Electrophysics, National Yang Ming Chiao Tung University, Hsinchu, ROC}

\author{Yi Yang}
\email{yiyang@nycu.edu.tw}
\affiliation{Department of Electrophysics, National Yang Ming Chiao Tung University, Hsinchu, ROC}
\affiliation{Center for Theoretical and Computational Physics, National Yang Ming Chiao Tung University, Hsinchu, ROC}
\affiliation{National Center for Theoretical Physics, ROC}

\begin{abstract}
We study the generalized entanglement entropy in the higher dimensional two-sided eternal black hole by double holography. By introducing an end-of-the-world ETW brane, which defines the time-dependent effective Hawking radiation region, we find a new type of Ryu-Takayanagi surface besides the Hartman-Maldacena surface and the island Ryu-Takayanagi surface known previously. We study the phase transition among the three Ryu-Takayanagi surfaces at different temperatures and obtain the phase diagram as well as the Page curve.
\end{abstract}

\maketitle

\section{Introduction}

The calculations of Hawking showed that black holes have thermodynamic properties like temperature and entropy. The idea is that a pair production that occurs near the horizon could lead to one particle to fall into the black hole and the other to escape to infinity, this leads to the well known Hawking radiation \cite{BF02345020,PhysRevD.13.191}. Since a black hole is formed from a pure state, this state should evolve in a way that it ends up in a final state that is also pure if the evolution were to obey unitarity. However, as the black hole evaporates completely, all that is left is thermal radiation so that the final state is in a mixed state. Apparently, the information that fell into the black hole vanishes. This leads to the black hole information paradox and violates unitarity which is one of the fundamental principle of quantum mechanics.

Don Page showed that if a black hole is formed from a pure state and evaporates unitarily then the von Neumann entropy of Hawking radiation should initially rise until the so-called Page time when it starts to drop down to zero as the black hole completely vanishes. This corresponds to the process where information can leak out from the black hole and is encoded in the Hawking radiation. However, this can only happen at late times past the Page time when the black hole has evaporated around half of its original state. This trend followed by the Hawking radiation is called the Page curve \cite{9306083}. Over the years, there were many proposals in addressing the black hole information paradox \cite{1207.3123, 1211.6767, 1301.4504, 1304.6483, 1306.0533, 1308.4209, 1310.6334, 1607.05141, 1712.04955, 1910.00972, 1911.06305, 2002.03543,2103.07477,2105.06579,2105.12737,2109.14618,2110.04233,2111.09551}. It was believed that the Page curve can only be obtained in the quantum gravity theory.

It has been shown by Bekenstein and Hawking that the entropy of a black hole is proportional to its horizon area which is the so-called Bekenstein-Hawking (BH) formula \cite{BF02757029, PhysRevD.7.2333}. This is a clear demonstration of the connection between a quantum-mechanical quantity and a geometric quantity. This connection was later generalized by Ryu and Takayanagi (RT) through the AdS/CFT correspondence, where they proposed a prescription connecting the entanglement entropy of a region $\mathcal{A}$ in a field theory to an area of a codimension two minimal surface $\mathcal{E}_\mathcal{A}$, called the RT surface, in its static dual bulk spacetime \cite{0603001,0605073}. This prescription was subsequently generalized by Hubeny, Rangamani and Takayanagi (HRT)  to the HRT surface in a time dependent bulk spacetime \cite{0705.0016}. The entanglement entropy calculated in this way is called the holographic entanglement entropy.

To include the quantum correction, the generalized entropy (fine-grained or von Neumann entropy) in a CFT was proposed in \cite{1307.2892}. Soon after, the holographic prescription for the generalized entropy was proposed by including the bulk entropy and finding a quantum extremal surface (QES) \cite{1408.3203,1904.08423}. Remarkably, based on the entanglement wedge reconstruction \cite{1601.05416}, the authors in \cite{1905.08255,1905.08762} showed that the Page curve can be obtained in the semiclassical calculation of the generalized entropy by holographic correspondence.

The QES in the context of a two-dimensional Jackiw-Teitelboim (JT) gravity theory coupled to non-gravitational conformal matter, i.e. the so called thermal bath, has been studied in \cite{1908.10996,1911.03402}. The bulk entropy was calculated by the double holography scheme. It was found that in order for the Hawking radiation to follow the Page curve for an evaporating black hole, an island should emerge along with a QES inside the horizon at late times to account for unitarity. While for the eternal black hole, it was shown that the QES could be outside the horizon \cite{1910.11077,2004.00598, 2004.01601, 2107.14746}.

The island formula has soon been extended to higher dimensional spacetime \cite{1910.12836,1911.09666,2004.05863,2005.08715,2006.02438,2006.04851,2006.10754,2006.11717,2006.16289,2007.06551,2007.15999,2010.00018,2010.00037,2010.04134,2010.16398,2012.03983,2012.05425,2012.07612,2101.05879,2101.10031,2101.12529,2102.01810,2102.02425,2103.15852,2103.16163,2104.00006,2104.02801,2104.07039,2104.00224,2104.13383,2105.00008,2105.01130,2105.08396,2105.09106,2106.07845,2106.10287,2106.11179,2106.12593,2107.00022,2107.01219,2107.03390}. Using the Randall-Sandrum (RS) brane scenario to study the QES in a higher dimensional thermofield double state has been discussed pioneeringly in \cite{1911.09666,2006.02438}. Later on, to combine both the RS brane scenario and Dvali-Gabadadze-Porrati (DGP) gravity \cite{0005016}, the intrinsic curvature of the branes is added in the action of the modified braneworld theory \cite{2006.04851}.

Several works which studied the Page curve more closely and the validity of the island formula can be found in \cite{2003.05448, 2003.11870, 2004.13857, 2004.14944, 2007.04877, 2007.10523, 2101.06867, 2103.14364, 2104.00052, 2105.12211, 2106.14738, 2107.03031, 2107.05218, 2107.07444, 2108.10144, 2110.07598}. A study which showed the possible existence of quantum extremal surfaces and entanglement wedges in flat space can be found in \cite{2005.02993}. Also, the calculation of a Page curve consistent with unitarity that relied on semiclassical calculations and did not need the island formulation can be found in \cite{2004.15010}. A review of the information loss paradox and its resolution using the island formulation can be found in \cite{2006.06872, 2012.05770,2107.09228}.

Later on, coupling the JT gravity theory to a thermal bath at a finite temperature has been studied in \cite{2007.11658,2102.00922,2107.10358,2109.01895,2109.03841}, where the doubly holographic bulk spacetime is a black hole instead of a pure AdS spacetime. However, coupling a higher dimensional gravity theory to a thermal bath at a finite temperature is still not clear.

On the other hand, boundary conformal field theory (BCFT) is a conformal field theory defined on a manifold with boundaries where suitable boundary conditions are imposed \cite{9505127}. Early studies of holographic dual of defect or interface CFT can be found in \cite{0011156,0111135}. The holographic dual of BCFT by including extra boundaries in the gravity dual was proposed in \cite{1105.5165,1108.5152,1205.1573}. In other words, a holographic construction of conformal field theories with boundaries can be established with some appropriate boundary conditions \cite{1701.04275,1701.07202}. The holographic entanglement entropy has been studied in \cite{1701.04275,1701.07202,1805.06117}. Using holographic BCFT to study the Page curve has been addressed in \cite{1910.12836,1911.09666,2006.16289}.

\begin{figure}[t]
	\centering
    \includegraphics[width=0.4\textwidth]{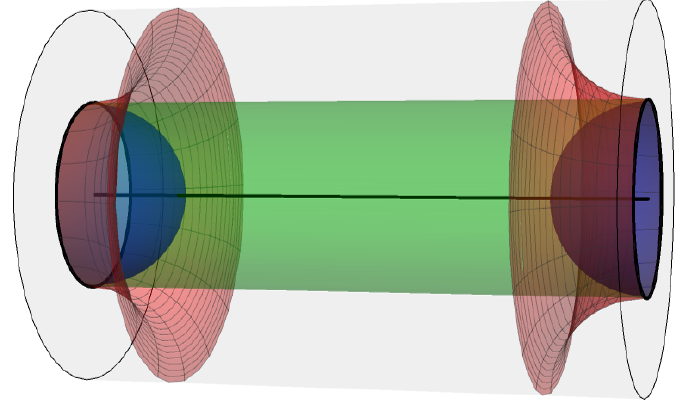}
	\caption{The three-dimensional illustration of the BCFT model where different RT surfaces are displayed. The green inner cylinder is the Hartman-Maldacena surface, the red hat-like surfaces are the boundary RT surfaces, and the blue hemispheres are the island RT surfaces. The boundary/ETW brane of the CFT is represented by the outer cylinder, while the Planck brane is the axis in the middle of the cylinder.}
	\label{BCFT3D}
\end{figure}

In the previous studies, the radiation region was considered to be the half infinite space, i.e. the thermal bath, and the entanglement entropy between the radiation region and the gravitational region including the black hole was computed. However, we cannot practically measure the full information in a half \textit{infinite} space. In fact, before a black hole couples to a thermal bath, the radiation from the black hole reflects back at the boundary of the gravitational region. The radiation starts to enter the radiation region when a thermal bath is coupled, and travels to infinity at the speed of light. The radiation front forms a moving surface in the thermal bath. The actual radiation region should be the time dependent finite region between the boundary and the moving surface.

In this paper, we use the modified braneworld theory to study the holographic entanglement entropy of a ($d+1$)-dimensional BCFT at finite temperatures, whose dual bulk spacetime is an asymptotically $\rm{AdS}_{d+2}$ black hole. We introduce an end-of-the-world (ETW) brane which is the hypersurface representing the  Hawking radiation front. This ETW brane defines a time-dependent effective radiation region, which supports a new type of RT surface instead of the two known ones, see the sketch in Fig.\ref{BCFT3D}. The two black circles on the two ends represent the entanglement surfaces. There are three possible RT surfaces that may appear. One is the two disconnected blue hemispheres that anchors on the entanglement surfaces. Another is the green inner cylinder which connects the entanglement surfaces directly. In addition, there is a third one, the two red hat-like surfaces, if there exists an ETW brane represented by the outer cylinder. This RT surface anchors on both the entanglement surfaces and the ETW brane. We examine the competition among the three RT surfaces that may appear in the course of the evolution of the Page curve by studying the phase transitions among them and the phase diagram.

This paper is organized as follows. In section \ref{Background and Setup}, we review the basics of holographic entanglement entropy and discuss our setup including the purification, double holography, and BCFT that will be used in our work. In section \ref{Holographic Entanglement Entropy of Hawking Radiation}, we calculate the holographic entanglement entropy for the three RT surfaces. In section \ref{Phase Transition and Page Curves}, we examine the phase transitions among the RT surfaces then discuss the phase diagram and the Page curve at different temperatures. We conclude our results in section \ref{Conclusion}.

\section{Background and Setup} \label{Background and Setup}

\subsection{Generalized Entanglement Entropy}

\begin{figure*}[t]
    \centering
    \subfloat[Sunset]{\includegraphics[width=0.3\textwidth]{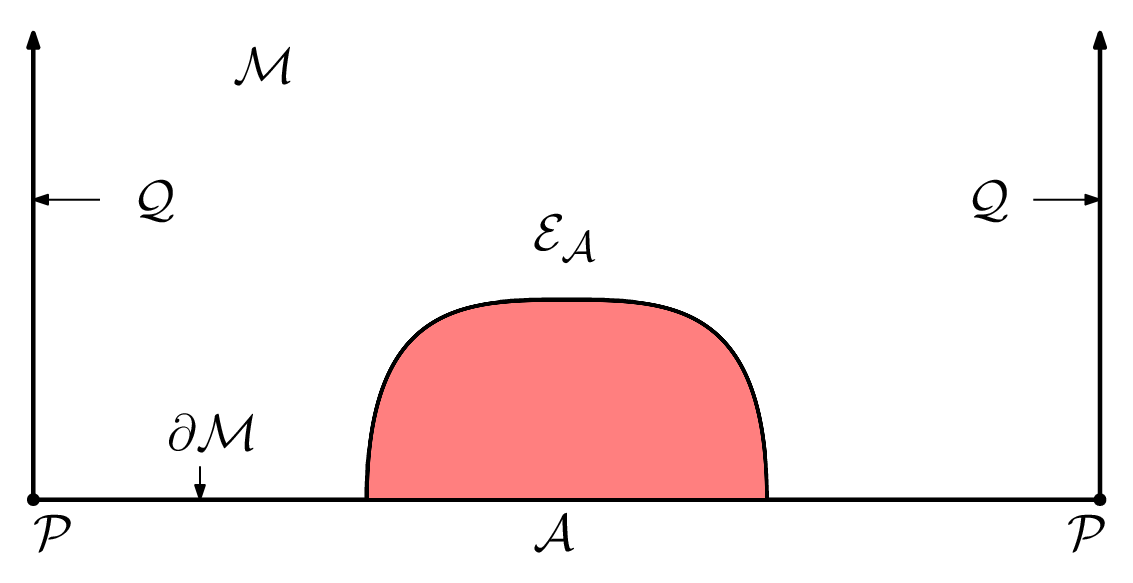}}
    \hspace{0.5cm}
    \subfloat[Rainbow]{\includegraphics[width=0.3\textwidth]{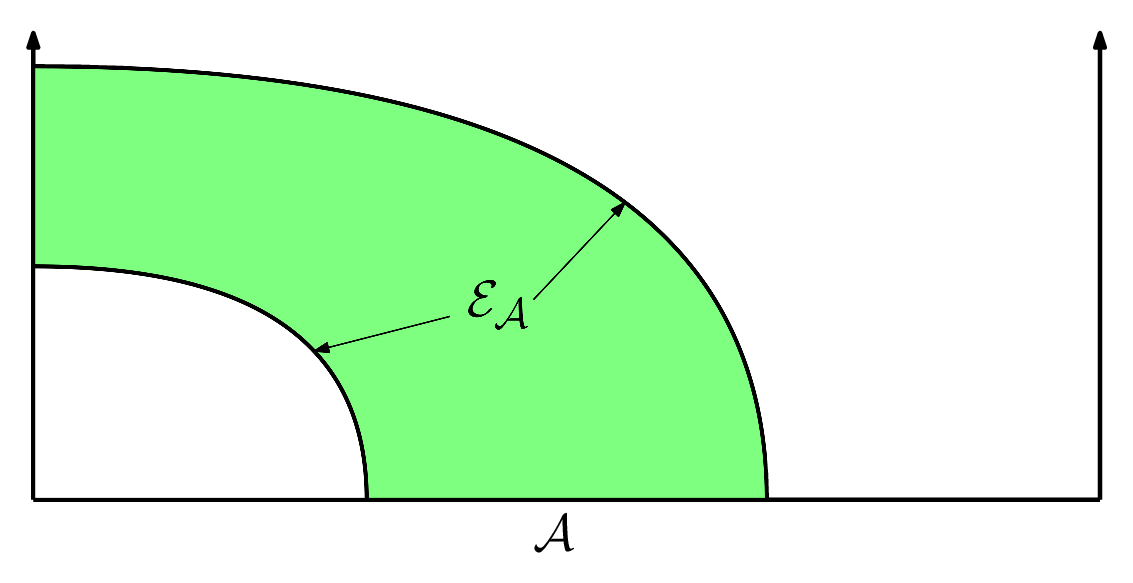}}
    \hspace{0.5cm}
    \subfloat[Sky]{\includegraphics[width=0.3\textwidth]{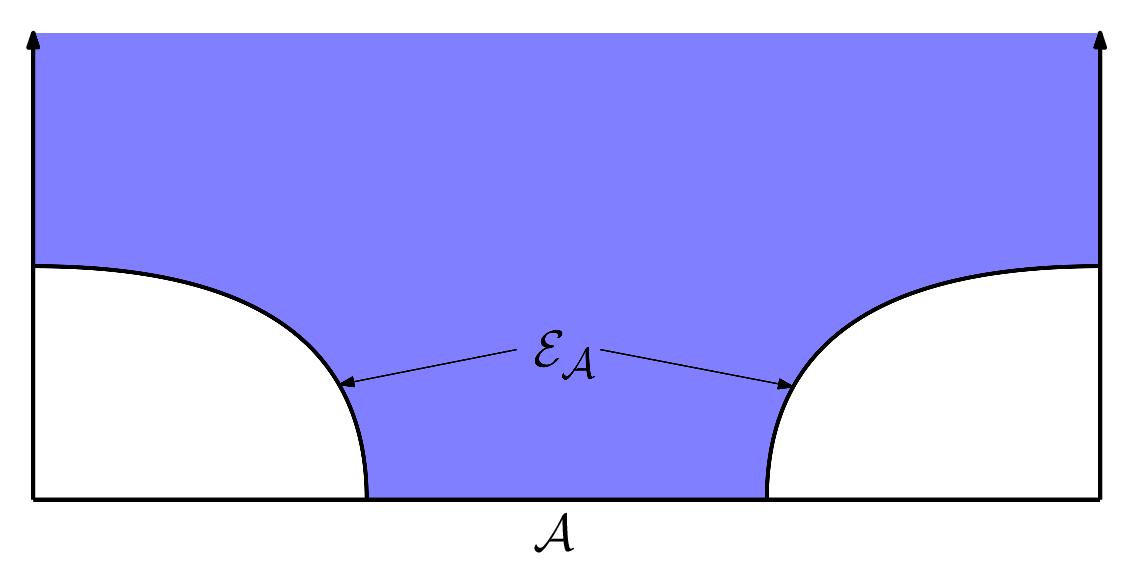}}
    \caption{Three possible RT surfaces $\mathcal{E}_\mathcal{A}$ of an entangling region $\mathcal{A}$ extending to the bulk spacetime $\mathcal{M}$ where the holographic dual gravity theory lives. The colored regions represent the entanglement wedges in different configurations. The definition of the notation can be found in Sec.\ref{Brief Review of Black Hole Solution in BCFT}}
    \label{RT}
\end{figure*}

Consider a Hilbert space $\mathcal{H}$ of a quantum field theory on a Cauchy time slice and divide $\mathcal{H}$ into two regions $\mathcal{A}$ and its complement $\mathcal{A}^c$. We denote the density matrix of the states in $\mathcal{H}$ by $\rho$. The entanglement entropy between the states in $\mathcal{A}$ and $\mathcal{A}^c$ is defined as,
\begin{equation}
  \mathcal{S}_\mathcal{\mathcal{A}} = - \rm{Tr} \left[ \rho_{\mathcal{\mathcal{A}}}\ln\rho_{\mathcal{A}} \right],
\end{equation}
where
\begin{equation}
  \rho_{\mathcal{\mathcal{A}}}=\rm{Tr}_{\mathcal{A}^c}\rho,
\end{equation}
is the reduced density matrix of the region $\mathcal{A}$.

Assuming the quantum field theory has a dual gravity theory living in a higher dimensional asymptotic AdS bulk spacetime $\mathcal{M}$ through the holographic correspondence \cite{9711200,9802150}, the entanglement entropy between $\mathcal{A}$ and $\mathcal{A}^c$ can be calculated by the Ryu-Takayanagi (RT) formula \cite{0603001,0705.0016},

\begin{equation}
  S^{\rm{class}}_{\mathcal{A}} = \underset{\mathcal{E}_\mathcal{A}}{\rm{min}} \frac{\text{Area}(\mathcal{E}_\mathcal{A})}{4G_N},
\end{equation}
where $G_N$ is the Newton constant and $\mathcal{E}_\mathcal{A}$ is a codimension two RT surface in the bulk spacetime $\mathcal{M}$ which is anchored on the boundary $\partial \mathcal{A}$ of the entangling region $\mathcal{A}$. It is important that the RT surface $\mathcal{E}_\mathcal{A}$ is homologous to $\mathcal{A}$. The presence of the boundaries $\mathcal{Q}$ in the BCFT spacetime allows the RT surface $\mathcal{E}_\mathcal{A}$ to reach the boundary $\mathcal{Q}$ as shown in Fig.\ref{RT}.

To include the quantum correction, we need to consider the generalized holographic entanglement entropy \cite{1307.2892},
\begin{equation} \label{GEE}
    S^{\rm{gen}}_{\mathcal{A}}(\mathcal{E}_\mathcal{A})= \frac{\text{Area}(\mathcal{E}_\mathcal{A})}{4G_N} + \mathcal{S}_{\rm{bulk}}(\mathcal{E}_\mathcal{A}),
\end{equation}
where $\mathcal{S}_{\rm{bulk}}(\mathcal{E}_\mathcal{A})$ is the bulk entropy between the entanglement wedge $\mathcal{R}_\mathcal{A}$ bounded by $\mathcal{A} \cup \mathcal{E}_\mathcal{A}$, and its complementary region in the bulk space.

The fine-grained entanglement entropy is defined as the minimal value of the generalized entanglement entropy,

\begin{equation} \label{Fine-Grained Entropy}
  \mathcal{S}_\mathcal{A} = \underset{X}{\rm{min}} \left[S^{\rm{gen}}_{\mathcal{A}}(X)\right] = \underset{X}{\rm{min}}  \left[\frac{\text{Area}(X)}{4G_N} + \mathcal{S}_{\rm{bulk}}(X)\right],
\end{equation}
where $X$ is the QES that minimizes the generalized entanglement entropy $S^{\rm{gen}}_{\mathcal{A}}(X)$.

\subsection{Black Hole Coupled to Thermal Bath} \label{Black Hole Coupled to Thermal Bath}

In the papers of $\rm AEM^4Z$ \cite{1905.08762,1908.10996}, a $2\rm d$ Jackiw-Teitelboim (JT) gravity on $\rm AdS_2$ is coupled to a matter $\rm CFT_2$. This 2d system is holographically dual to a 1d quantum mechanical system. If this $2\rm d$ theory containing a black hole is coupled to a bath which consists of the same $\rm CFT_2$ but now living on a flat space, then the black hole is permitted to evaporate into the bath.

\begin{figure*}[t]
	\centering
	\includegraphics[scale=0.75]{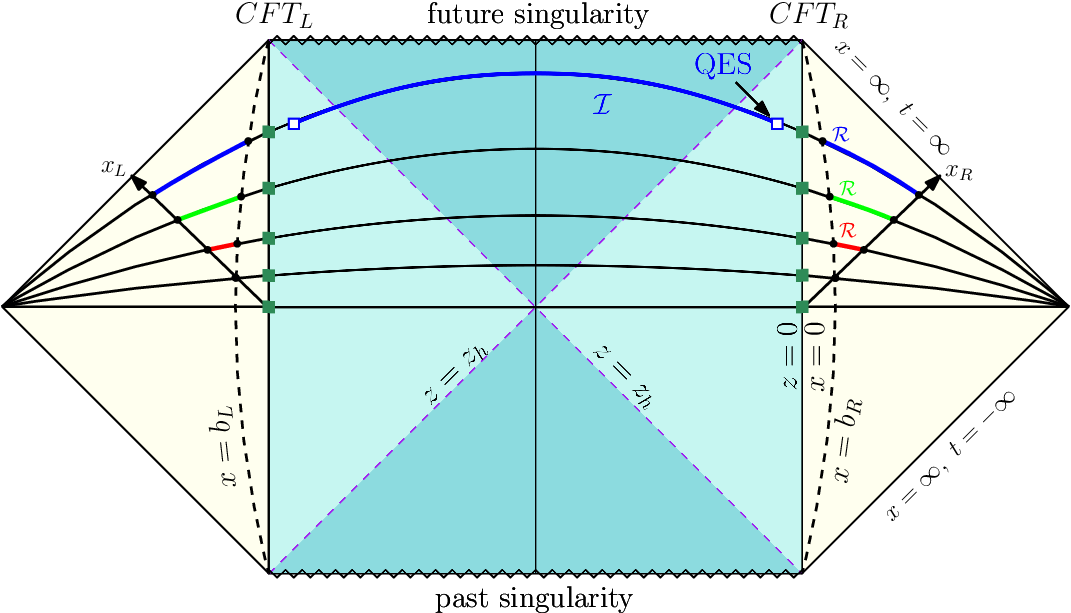}
	\caption{The Penrose diagram of the two-sided eternal black hole. The violet dashed lines represent the event horizon of the black hole ($z = z_h$), which separates the black hole into interior (darker blue) and exterior (lighter blue) regions. Two copies of the CFT live on the two boundaries of the eternal black hole. Two copies of the bath (light yellow) are coupled to the boundaries of the black hole. The vertical dashed curves are the cutoff at $x = b_{L,R}$. The black arrow represents the time-dependent ETW brane $\mathcal{Q}_E$. The horizontal curves are Cauchy slices at different boundary times with the red, green, and blue fragments representing the entanglement regions of the Hawking radiation.}
	\label{Penrose 1}
\end{figure*}

To study the Hawking radiation for an evaporating black hole, one has to deal with a time-dependent spacetime. Alternatively, one can consider a thermofield double state (TFD) that is dual to a two-sided eternal black hole \cite{1910.11077}. The Penrose diagram of the eternal black hole system is shown in Fig.\ref{Penrose 1}. There are two copies of the CFT, $\rm{CFT}_L$ and $\rm{CFT}_R$, living on the boundaries of two copies of the exterior region (shaded in lighter blue color), which are connected by a wormhole (shaded in darker blue color). The bifurcation dashed lines are the event horizons of the eternal black hole.

The system is time-invariant as time evolves forward on one side while backwards on the other side. However, one can consider time evolving forward on both sides to introduce the time dependence in this system \cite{1303.1080}.

In addition, in order to allow the eternal black hole to radiate, we couple two copies of an auxiliary thermal bath \cite{2104.00183} (shaded in light yellow color) at the boundaries of the two exterior regions. The two copies of the thermal bath are in flat spacetime and have the same finite temperature $T$ with the black hole. The eternal black hole together with the two copies of the thermal bath comprises a pure system.

We are going to calculate the entanglement entropy of the CFTs on the boundaries of the exterior regions using the generalized entanglement entropy in eq.(\ref{Fine-Grained Entropy}). To do so, we introduce a cutoff at $x=b_{L,R}$ which is slightly inside the left or right bath regions. The cutoff separates the whole system into two regions: the gravitational region that includes the black hole with its boundaries, and the radiation region where the Hawking radiation escapes. The entanglement region is thus the surface at the cutoff $x=b_{L,R}$.

Through the holographic correspondence, the entanglement entropy of the CFTs can be obtained by finding a minimal area surface in the bulk gravity spacetime that is homologous to the entanglement region. Classically, the minimal surface is the union of the two horizons, i.e. the classical RT surface, which is a time independent constant. The classical RT surface gives the coarse-grained entropy that sets the upper bound of the entanglement entropy of the Hawking radiation.

Including quantum corrections, the QES is the surface that minimizes the generalized entanglement entropy eq.(\ref{Fine-Grained Entropy}). The location of the QES on a Cauchy slice is generically different at different slices, so the fine-grained entropy of the eternal black hole is time-dependent. Furthermore, there could be more than one QES that minimize the entanglement entropy locally. The dominant QES should be the one which globally minimizes the entanglement entropy. This implies the possible phase transitions between different QES.

Having the basic picture in mind, let us now consider the radiation region in more detail. We couple the bath to the eternal black hole at a certain time $t=t_h$. We assign the bath to have the same temperature as that in the eternal black hole. As the Hawking radiation from the black hole escapes to the thermal bath, there is the same amount of energy that goes back to the black hole from the bath so that the whole system is in thermodynamic equilibrium.

To decode the information inside the black hole, we need to measure the Hawking radiation from the black hole travelling to the thermal bath. Once the Hawking radiation has entered the bath, it then travels along the outgoing null surface at the speed of light as shown in Fig.\ref{Penrose 1}. At time $t=t_b$, the earliest Hawking radiation reaches the cutoff at $b_{L,R} = c(t_b-t_h)$ and enters into the radiation region. We then start to measure the Hawking radiation from the time $t_b$ and set it as our initial time.

At a later time $t>t_b$, the earliest radiation will reach a surface at
\begin{equation}
 x_{L,R}=b_{L,R} + c(t-t_b) = c(t-t_h),
\end{equation}
which behaves as an ETW brane\footnote{We will set $c=1$ in the following of the paper.}. Since there is no radiation beyond the ETW brane, the effective radiation region is just the space-like region between the cutoff and the ETW brane, namely $[b_L,x_L] \cup [b_R,x_R]$, which increases with time.

As we have mentioned, the whole system including both the eternal black hole and the thermal bath is a pure system, so that the bulk entanglement entropy in eq.(\ref{Fine-Grained Entropy}) equals the entanglement entropy of the Hawking radiation in the effective radiation region $\mathcal{R}$ as shown in the Fig.\ref{Penrose 1}. The entanglement entropy of the radiation region $\mathcal{R}$ can be calculated using the island formula \cite{1908.10996, 2006.04851},
\begin{equation}
  \mathcal{S}_{EE}(\mathcal{R}) = \min \left\{ \text{ext} \left( S (\mathcal{R} \cup \mathcal{I}) + \frac{\text{Area}(\partial \mathcal{I})}{4 G_N^{(d+1)}} \right) \right\},
\end{equation}
It can be shown that, at early times, the entanglement entropy of the effective radiation region is always dominated by the vanishing island $\mathcal{I}=\emptyset$ with $\text{Area}(\partial \mathcal{I})=0$. The bulk entropy $S(\mathcal{R}\cup \emptyset)$ corresponding to the vanishing island, monotonically increases from zero and will eventually exceed the coarse grained entropy implying the information paradox.

Remarkably, there exists a non-trivial QES outside of the horizon for the eternal black hole \cite{1303.1080} that implies a non-vanishing island $\mathcal{I}$ between the two non-trivial QES. The bulk entropy corresponding to this non-vanishing island equals the entanglement entropy of the union of the radiation region with the island and will dominate at late times as shown in the Fig.\ref{Penrose 1}.

Since the QES is outside of the horizon, the generalized entanglement entropy for the non-vanishing island is a constant. At a later time, e.g. $t=t_I$ in Fig.\ref{Penrose 1}, the non-vanishing island case will dominate the system. The time at which the phase transition between the vanishing and non-vanishing island takes place is called the Page time. The phase transition of the generalized entanglement entropy leads to the well known Page curve for the eternal black hole.

The eternal black hole in 2-dimensions has been studied in \cite{1910.11077}. Extension to higher dimensions was previously considered in \cite{1911.09666,2006.02438,2006.04851}. The main difficulty in obtaining the generalized entanglement entropy is calculating the bulk entropy in curved spacetime. In 2-dimensions, the replica trick is used to compute the entanglement entropy through the path integral approach. However, this trick is difficult to be generalized to higher dimensions.

In section \ref{Double Holography}, we will use the idea of double holography to calculate the bulk entanglement entropy \cite{1908.10996}. It has been shown that combining double holography with BCFT is a powerful method to calculate the bulk entanglement entropy and can be generalized to higher dimensions \cite{2010.00018}.

\begin{figure*}[t]
	\centering
	\subfloat[boundary perspective]{\includegraphics[scale=0.8]{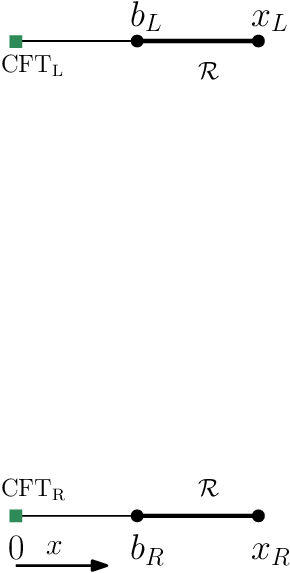}}
    \label{boundary perspective}
    \hspace{1cm}
    \subfloat[brane perspective]{\includegraphics[scale=0.8]{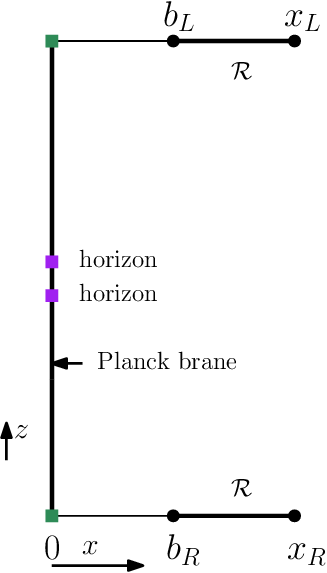}}
    \label{brane perspective}
    \hspace{1cm}
    \subfloat[bulk perspective]{\includegraphics[scale=0.8]{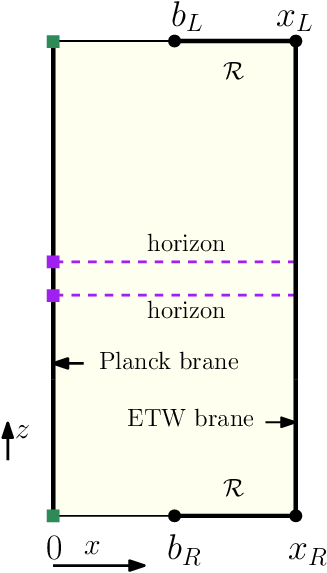}}
    \label{bulk perspective}
	\caption{Three different perspectives of the gravity system, Hawking radiation system, and the holographic dual of these systems.}
	\label{Three-Perspectives}
\end{figure*}

\subsection{Double Holography}\label{Double Holography}

To calculate the bulk entanglement entropy in JT gravity, the authors of \cite{1908.10996,1911.03402} introduced a $3\rm d$ spacetime that is holographically dual to the $2\rm d$ matter CFTs and the baths, i.e. the double holographic duality. The $3\rm d$ double holographic bulk metric is locally an $\rm AdS_3$ spacetime with a boundary on which the $2\rm d$ theory is living. This is similar to the RS model, where the boundary is called the Planck brane \cite{9905221,9906064,0011156}.

Extending the RS brane scenario to a higher dimensional thermofield double state has been discussed pioneeringly in \cite{1911.09666,2006.02438}. Later on in \cite{2006.04851}, the intrinsic curvature term of the branes is added in the action that combines the features of both the RS brane scenario and the DGP gravity \cite{0005016}. 

As shown in \cite{2006.02438}, there are three equivalent perspectives to describe the system.
In the \textit{boundary perspective}, a pair of $(d+1)$-dimensional CFTs live on a region $[0,x_{L,R}]$ with a $d$-dimension CFT living on its boundary at $x=0$ indicated by the green square dots as shown in Fig.\ref{Three-Perspectives}(a), where $x=b_{L,R}$ is the cutoff and the other boundary is at $x=x_{L,R}$. The effective Hawking radiation region is $[b_L,x_L] \cup [b_R, x_R]$.

The \textit{brane perspective} can be realized by using holographic correspondence to replace the $\rm CFT_{d}$ by a Planck brane on which an asymptotically $\rm AdS_{d+1}$ eternal black hole lives, as shown in Fig.\ref{Three-Perspectives}(b). We assign the radial direction of the eternal black hole as $z$ with the $\rm CFT_{d}$ sitting at $z=0$. The horizons of the eternal black hole are indicated by the two purple square dots. As argued in \cite{0011156,0105132,0109017,0106261,0112166,0201261,0207003,0608089,2006.02438}, the transparent boundary condition between the gravitational and the radiation regions imply that the stress tensor in $\rm CFT_{d}$ is not conserved and will generate an anomalous dimension, so that the dual graviton gets a mass. However, if the central charge of the $\rm CFT_{d}$ is large, the effect from the mass can be negligible.

In the \textit{bulk perspective}, we use the holographic correspondence again to replace the $\rm CFT_{d+1}$ by an asymptotically $\rm AdS_{d+2}$ black hole bulk spacetime. This geometry can be properly described by a holographic BCFT setup, as shown in Fig.\ref{Three-Perspectives}(c). The $(d+1)$-dimensional BCFT is dual to an $\rm AdS_{d+2}$ black hole bulk spacetime. The boundary $x=0$ extends to the bulk spacetime as a Plank brane and the other boundary $x=x_{L,R}$ extends to the bulk spacetime as an ETW brane. The embedding of the branes into the bulk spacetime is determined by solving the holographic BCFT system. The purple dashed lines represent the two horizons in the bulk black hole. In the bulk perspective, the bulk entanglement entropy in eq.\eqref{GEE} can be calculated by a \textit{classical} RT surface in the bulk spacetime anchored on the entangling surface at $x=b_{L,R}$.


In summary, our setup consists of two boundaries. One is the boundary on the left at $x=0$ which corresponds to the Planck brane, the other is the boundary on the right at $x=x_{L,R}$ which corresponds to the ETW brane. The effective radiation region is $[b_L, x_L] \cup [b_R, x_R]$.

\subsection{Brief Review of Black Hole Solution in BCFT} \label{Brief Review of Black Hole Solution in BCFT}

In this subsection, we will briefly review a black hole solution in the holographic BCFT that we will use in this work.

The $(d+2)$-dimensional bulk spacetime $\mathcal{M}$ is holographically dual to a $(d+1)$-dimensional CFT defined on the  conformal boundary $\partial \mathcal{M}$. $\partial \mathcal{M}$ has a $d$-dimensional boundary $\mathcal{P}$, which has a $(d+1)$-dimensional hypersurface dual $\mathcal{Q}$ in $\mathcal{M}$ anchored at $\mathcal{P}$, see Fig.\ref{RT}.

The action of the gravitational theory in the bulk is,
\begin{equation}
  S = \mathcal{S}_{\mathcal{M}} + S_{GHY} + S_{\mathcal{Q}} + S_{\partial \mathcal{M}} + S_{\mathcal{P}},
\end{equation}
where
\begin{align}
  S_{\mathcal{M}} & = \int_{\mathcal{M}} \sqrt{-g} (R - 2\Lambda_{\mathcal{M}}), \\
  S_{\mathcal{Q}} & = \int_{\mathcal{Q}} \sqrt{-h} (R_{\mathcal{Q}} - 2\Lambda_{\mathcal{Q}} + 2 K), \\
  S_{\partial \mathcal{M}} & = 2 \int_{\partial \mathcal{M}} \sqrt{-\gamma} K', \\
  S_{\mathcal{P}} & = 2 \int_{\mathcal{P}} \sqrt{-\sigma} \theta.
\end{align}
$S_{\mathcal{M}}$ is the Einstein–Hilbert action of $\mathcal{M}$ with $R$ and $\Lambda_{\mathcal{M}}$ being the intrinsic Ricci curvature and the cosmological constant of $\mathcal{M}$ respectively. $S_{GHY}$ is the Gibbons-Hawking-York boundary term. $S_{\mathcal{Q}}$ is the action of the $(d+1)$-dimensional hypersurface $\mathcal{Q}$ with the induced metric $h_{ab} = g_{ab} - n_{a}^{\mathcal{Q}} n_{b}^{\mathcal{Q}}$, where $n^{\mathcal{Q}}$ is the unit normal vector of $\mathcal{Q}$, and $R_{\mathcal{Q}}$, $\Lambda_{\mathcal{Q}}$, $K$ being the intrinsic Ricci curvature, the cosmological constant, and the trace of the extrinsic curvature of $\mathcal{Q}$ in $\mathcal{M}$ respectively. $S_{\partial \mathcal{M}}$ is the action of the conformal boundary $\partial \mathcal{M}$ with the induced metric $\gamma_{ab} = g_{ab} - n_{a}^{\partial \mathcal{M}} n_{b}^{\partial \mathcal{M}}$, where $n^{\partial \mathcal{M}}$ is the unit normal vector of $\partial \mathcal{M}$, and $K'$ is the trace of the extrinsic curvature of $\partial \mathcal{M}$. $S_{\mathcal{P}}$ is the boundary term of $\mathcal{Q}$ and $\partial \mathcal{M}$ with $\sigma_{ab}$ being the metric of $\mathcal{P}$ and $\theta = \cos^{-1} \left( n^{\mathcal{Q}} \cdot n^{\partial \mathcal{M}} \right)$ being the supplementary angle between $\mathcal{Q}$ and $\partial \mathcal{M}$, which makes a well-defined variational principle on $\mathcal{P}$.

Comparing the above BCFT system with the Penrose diagram in Fig.\ref{Penrose 1}, $\mathcal{Q}$ represents the Planck brane containing the eternal black hole (blue region), $\partial \mathcal{M}$ represents the thermal baths (light yellow region), and $\mathcal{P}$ is the boundary of the eternal black hole at $x=0$ where the CFTs live.

A simple asymptotic $AdS_{d+2}$ black hole solution of the above system has been obtained in \cite{1805.06117},
\begin{equation} \label{AdS BH metric}
  ds_{\mathcal{M}}^2 = \frac{l_{AdS}^2}{z^2} \Bigg[ -f(z) dt^2 + \frac{dz^2}{f(z)} + dx^2 + \sum_{i=1}^{d-1} (dx_i)^2 \Bigg],
\end{equation}
where $l_{AdS}$ is the AdS radius and
\begin{equation}
  f(z) = 1 - \frac{z^{d+1}}{z_h^{d+1}},
\end{equation}
with $z_h$ the horizon of the black hole. The $(d+1)$-dimensional conformal boundary $\partial \mathcal{M}$ is at $z = 0$. The temperature of the BCFT is given by the Bekenstein-Hawking temperature of the black hole,
\begin{equation} \label{Temperature}
  T = \frac{d+1}{4\pi z_h}.
\end{equation}
By varying $S_{\mathcal{Q}}$ with $h^{ab}$ we get the equations of motion for $\mathcal{Q}$,
\begin{equation}
  R_{\mathcal{Q} ab} + 2 K_{ab} - \left( \frac{1}{2} R_{\mathcal{Q}} + K - \Lambda_{\mathcal{Q}} \right) h_{ab} = 0,\label{Ncondition}
\end{equation}
which is the Neumann boundary condition proposed by Takayanagi in \cite{1105.5165}. However, the condition  (\ref{Ncondition}) is too strong which gives more constraint equations than the degrees of freedom. In \cite{1701.04275, 1701.07202}, Chu et al proposed the following mixed boundary condition,
\begin{equation} \label{mixed BC}
  (d-1) (R_{\mathcal{Q}} + 2K) - 2(d+1) \Lambda_{\mathcal{Q}} = 0.
\end{equation}

\begin{figure*}[t]
  \centering
  \includegraphics[scale=0.7]{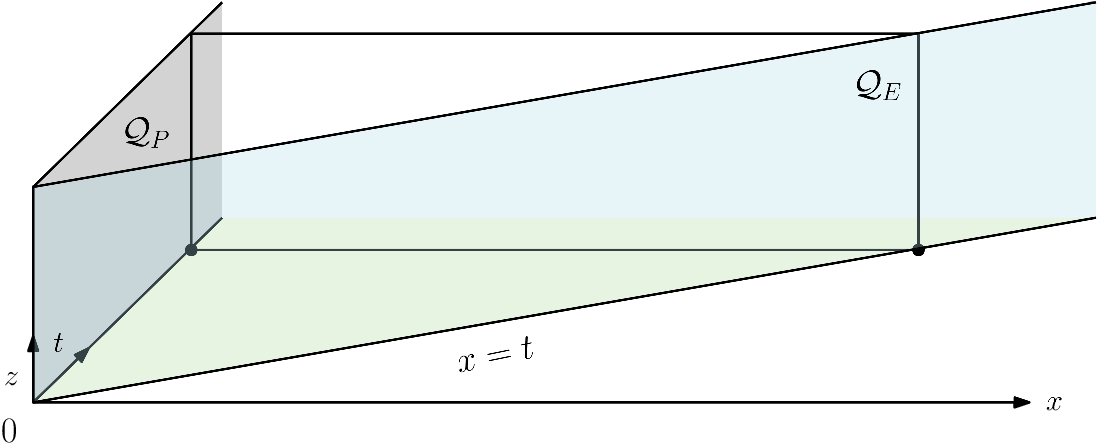}
  \caption{The embedding of the Planck brane $\mathcal{Q}_P$ and the ETW brane $\mathcal{Q}_E$ where $\mathcal{Q}_P$ is described by $x=0$, while the ETW brane $\mathcal{Q}_E$ is described by $t=x$.}
  \label{QPQE}
\end{figure*}

In our double holographic setup, we will use two simple solutions of the $(d+1)$-dimensional hypersurface, namely the Planck brane $\mathcal{Q}_P$ and the ETW brane $\mathcal{Q}_E
$. The Planck brane $\mathcal{Q}_P$ is time independent and has an embedding equation $x=0$, while the ETW brane $\mathcal{Q}_E$ is time-dependent and described by the equation $t=x$, these are shown in Fig.\ref{QPQE}. It is straightforward to show that the intrinsic curvature, trace of the extrinsic curvature, and the cosmological constant for the Planck brane $\mathcal{Q}_P$ and the ETW brane $\mathcal{Q}_E$ are,
\begin{equation}
  R_{\mathcal{Q}_P} = -\frac{d(d+1)}{l_{AdS}^2}, ~ K = 0, ~ \Lambda_{\mathcal{Q}_P} = -\frac{d(d-1)}{2 l_{AdS}^2},
\end{equation}
and
\begin{equation}
 R_{\mathcal{Q}_E} = -\frac{d(d+1)}{3l^2_{AdS}}, ~ K= 0, ~ \Lambda_{\mathcal{Q}_E} = -\frac{d(d-1)}{6l^2_{AdS}},
\end{equation}
which satisfy the mixed boundary condition eq.(\ref{mixed BC}).

We would like to remark that the ETW brane $\mathcal{Q}_E$ has the characteristics of a null cone only at $z=0$, while away from $z=0$ this is not necessarily true. Nevertheless, a sketch of the BCFT setup is shown in Fig.\ref{QPQE}.

\section{Holographic Entanglement Entropy of Hawking Radiation} \label{Holographic Entanglement Entropy of Hawking Radiation}

As we have explained in section \ref{Double Holography}, we are going to calculate the entanglement entropy with the effective radiation region $[b_L,x_L] \cup [b_R, x_R]$ as the entanglement region. By holographic correspondence, it is proportional to the area of the RT surface in the doubly holographic bulk spacetime. The RT surface is anchored at the entanglement surface $x=b_{L,R}$ and is homologous to the effective radiation region $[b_L,x_L] \cup [b_R, x_R]$. At the leading order, we only need to consider the classical RT surface. The area of the surface is given by the Nambu-Goto string action,

\begin{equation} \label{action}
S = \int d^{d+1} \xi \mathcal{L} =\int d^{d+1} \xi \sqrt{\det G_{ab}},
\end{equation}
where the induced metric is defined by
\begin{equation}
G_{ab}=g_{\mu\nu} \partial_a X^\mu \partial_b X^\nu,
\end{equation}
and the bulk metric $g_{\mu\nu}$ is given in eq.(\ref{AdS BH metric}).

Generically, there could be more than one locally minimum surface. In the following of this section, we will find that there are three minimum surfaces in our system. One is the \textit{Hartman-Maldacena surface} (HM), which penetrates the horizon and connects the cutoff at $x=b_{L,R}$. Another one is the \textit{boundary RT surface} (BRT), which is anchored not only on the cutoffs at $x=b_{L,R}$ but also on the ETW brane $\mathcal{Q}_E$. The third one is the \textit{island RT surface} (IRT) that is anchored on the cutoff $x=b_{L,R}$ and on the Planck brane $\mathcal{Q}_P$ at the QES $z=z_{\rm QES}$.

\begin{figure*}[t]
  \centering
  \subfloat[]{\includegraphics[scale=0.4]{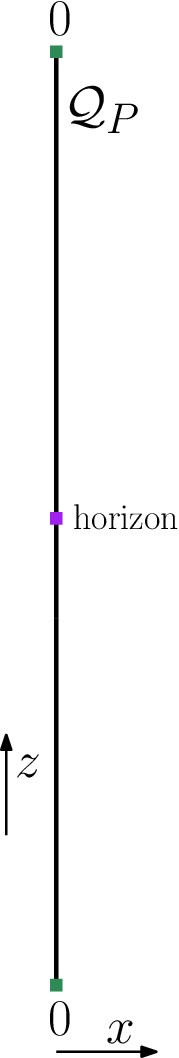}}
  \hfill
  \subfloat[]{\includegraphics[scale=0.4]{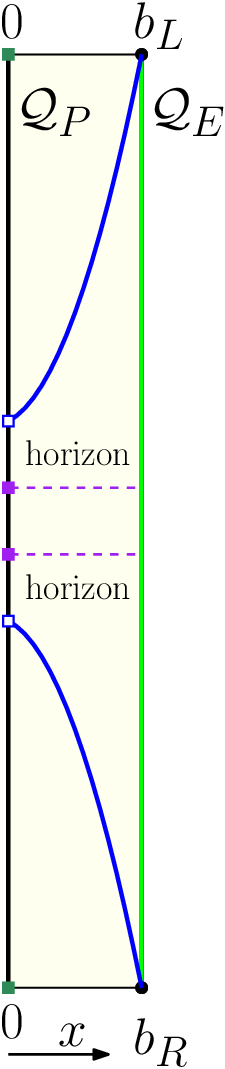}}
  \hfill
  \subfloat[]{\includegraphics[scale=0.4]{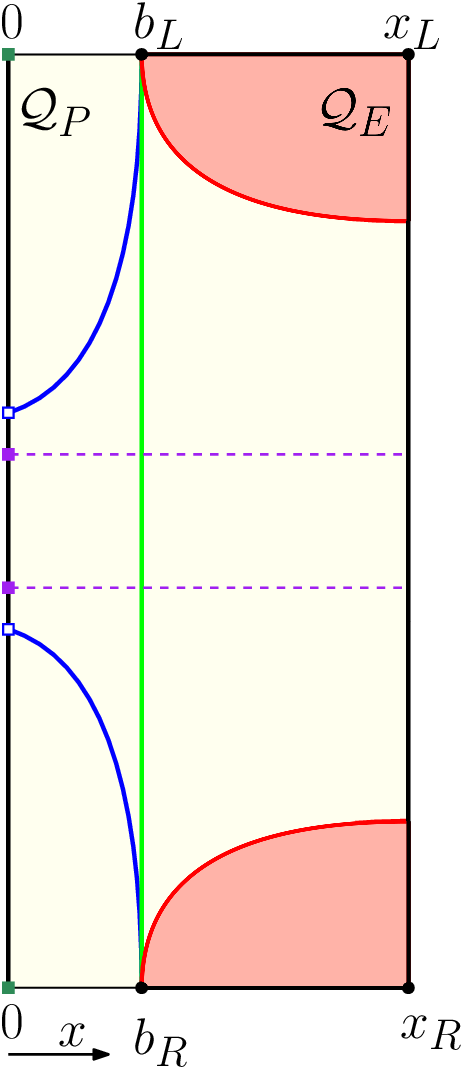}}
  \hfill
  \subfloat[]{\includegraphics[scale=0.4]{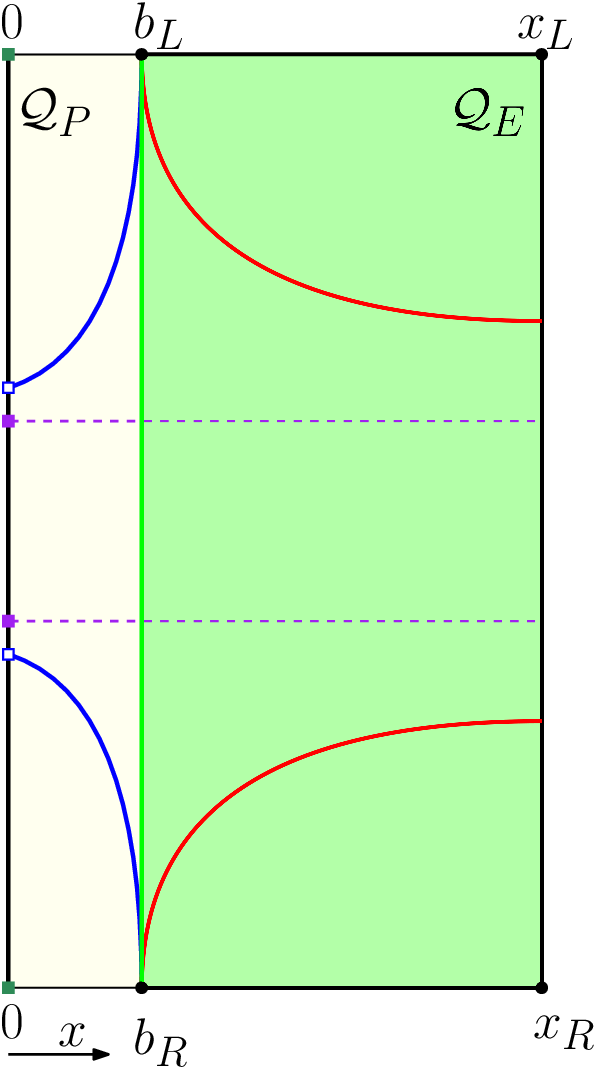}}
  \hfill
  \subfloat[]{\includegraphics[scale=0.4]{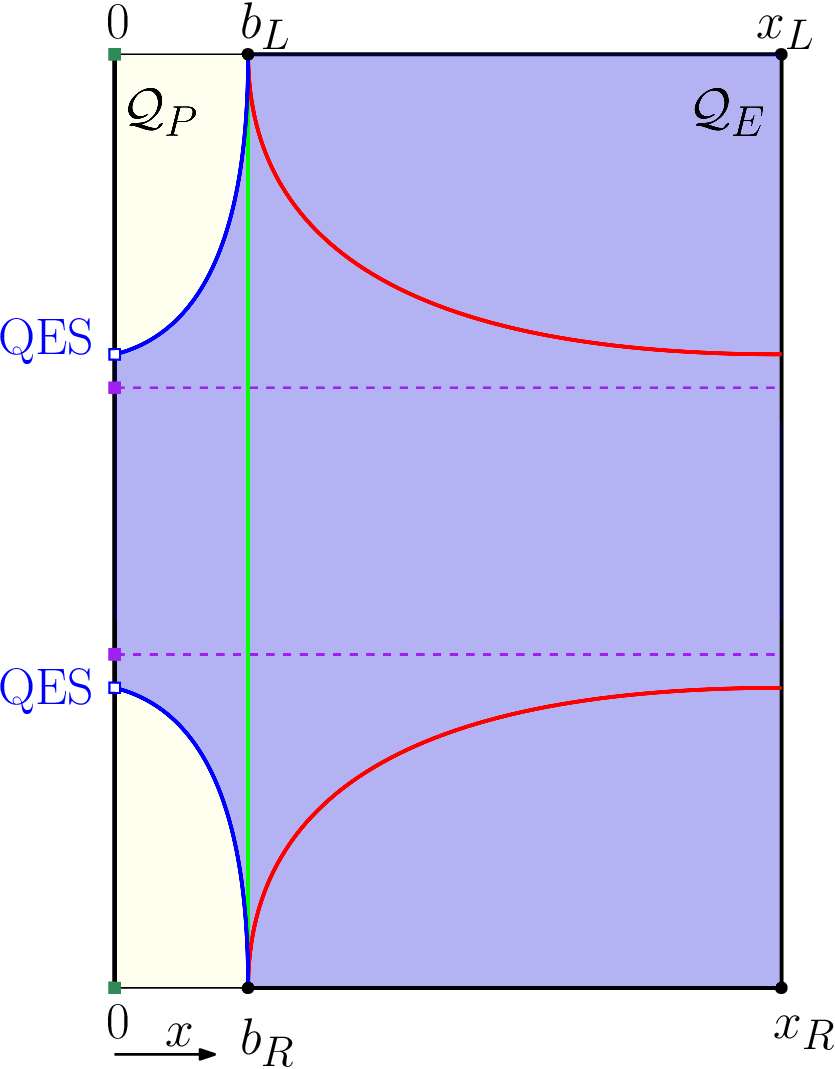}}
  \caption{The time evolution of the RT surfaces in the BCFT system. The red/green/blue line represents the BRT/HM/IRT surface. The color of the shaded region represents the entanglement wedge whose entanglement entropy is dominant at that time. $\mathcal{Q}_P$ and $\mathcal{Q}_E$ label the Planck brane and the ETW brane, respectively.}
  \label{RT surfaces}
\end{figure*}

The RT surfaces at different times are shown in Fig.\ref{RT surfaces}. The red/green/blue line represents the BRT/HM/IRT surface. At time $t=t_h$, as shown in  Fig.\ref{RT surfaces}(a), the Hawking radiation from the black hole reaches the boundary $z=0$ and enters into the thermal bath at $x=0$. At the initial time $t=t_b$, as shown in  Fig.\ref{RT surfaces}(b), the Hawking radiation reaches the cutoff $x=b_{L,R}$ and enters into the effective radiation region. Since the ETW brane locates at $x=b_{L,R}$ at the initial time $t=t_b$, the BRT surface vanishes, while only the HM and the IRT surfaces are present as shown in Fig.\ref{RT surfaces}(b).

The BRT surface dominates the system at the early time with the red region being the entanglement wedge as shown in Fig.\ref{RT surfaces}(c). As time evolves, the stretching of the interior of the black hole causes the growth of the HM surface. However, at the same time, the increase of the location of the ETW brane allows the BRT surface to increase faster than the HM surface. At a certain time, the BRT surface will exceed the HM surface and the latter becomes dominant with the green region being the entanglement wedge as shown in Fig.\ref{RT surfaces}(d).

Finally, after a critical time, both the BRT and HM surfaces will exceed the IRT surface so that the IRT becomes dominant with the blue region being the entanglement wedge as shown in Fig.\ref{RT surfaces}(e). When the IRT surface dominates, the entanglement wedge includes the part between the QESs on the Planck brane, i.e. the island. Since the QES is outside of the horizon in the eternal black hole, the island contains the whole interior of the black hole. Based on the entanglement wedge reconstruction, after the critical time, we are able to reconstruct all the information inside the black hole from the data observed in the effective radiation region. The critical time is called the Page time $t_P$.

In the following calculations, we are going to denote the $(d-1)$-dimensional volume as,
\begin{equation}
  \int d^{d-1} \mathbf{x} = V_{d-1}.
\end{equation}

\subsection{Hartman-Maldacena Surface} \label{Hartman-Maldacena Surface}

The HM surface is a locally minimum surface penetrating the horizon and connecting the two cutoffs at $x=b_{L,R}$. To describe the interior region behind the horizon, we make the following coordinate transformation,

\begin{equation} \label{HM coordinate transformation}
  t = v + \int \frac{dz}{f(z)} ~ \Rightarrow dt = dv + \frac{dz}{f(z)}.
\end{equation}
The metric eq.\eqref{AdS BH metric} then becomes,
\begin{equation}
  ds^2 = \frac{l^2_{AdS}}{z^2} \left[ -f(z) dv^2 - 2dvdz + dx^2 + \sum_{i=1}^{d-1} dx_i^2 \right],
\end{equation}
where the HM surface is described by $x=b_{L,R}$ and $v = v(z)$.

\begin{figure}
  \centering
  \includegraphics[width=0.4\textwidth]{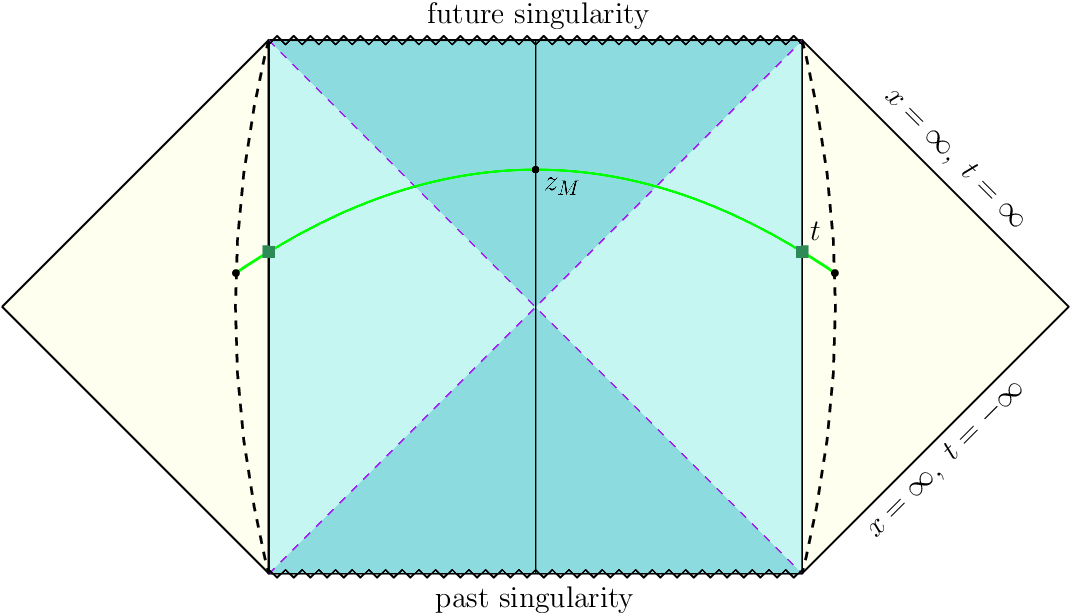}
  \caption{The Hartman-Maldacena Surface at time $t$.}
  \label{HM-RT}
\end{figure}

The string action eq.(\ref{action}) on the $t-z$ plane (or the $v-z$ plane) after the coordinate transformation eq.(\ref{HM coordinate transformation}), can be calculated as,
\begin{equation} \label{SHMdef}
  \mathcal{S}_{\rm HM} = \frac{l_{AdS}^d V_{d-1}}{4 G_N^{(d+2)}} \int dz \frac{1}{z^d} \sqrt{-\left(\frac{dv}{dz}\right) \left( f(z) \left(\frac{dv}{dz}\right) + 2 \right)}.
\end{equation}
The conjugate momentum of $v$ is,
\begin{equation} \label{conjugate momentum}
  \frac{\partial \mathcal{L}}{\partial v'} = C_v = \frac{f(z) v' + 1}{z^d \sqrt{-v' \left( f(z) v' + 2 \right)}},
\end{equation}
where $v' = {dv}/{dz}$.
From eq.(\ref{conjugate momentum}) we can solve for $v'$ as,
\begin{equation} \label{dvdz}
  v' = \frac{1}{f(z)} \left[ -1 \pm \sqrt{\frac{C_v^2 z^{2d}}{f(z) + C_v^2 z^{2d}}} \right].
\end{equation}
We choose the solution with the negative square root and integrate it on $z$ to get,
\begin{equation} \label{vinteg}
    v = \int_0^{z_M} \frac{dz}{f(z)} \left[ -1 - \sqrt{\frac{C_v^2 z^{2d}}{f(z) + C_v^2 z^{2d}}} \right],
\end{equation}
where $z_M$ is the turning point of the HM surface with the boundary time $t$ as shown in Fig.\ref{HM-RT}. Substituting eq.(\ref{vinteg}) back into eq.(\ref{HM coordinate transformation}), we obtain the expression of the boundary time,
\begin{equation} \label{tinteg}
    t = -\int_{0}^{z_M} \frac{C_v z^{d} dz}{f(z) \sqrt{f(z) + C_v^2 z^{2d}}}.
\end{equation}
The turning point of the HM surface can be found by imposing a boundary condition at $z = z_M$,
\begin{equation}
  \left. \frac{dz}{dv} \right|_{z = z_M} = f(z_M) \sqrt{\frac{f(z_M) + C_v^2 z_M^{2d}}{-f(z_M)}} = 0,
\end{equation}
which leads to the following expression of the conjugate momentum $C_v$ in terms of $z_M$,
\begin{equation} \label{Cv}
 C_v^2 = -\frac{f(z_M)}{z_M^{2d}}.
\end{equation}
As we have mentioned above, we will use
\begin{equation} \label{tHMh}
  t_h = - \int_{0}^{z_h} \frac{\sqrt{f(z_h)} z^d dz}{f(z) \sqrt{f(z_h) z^{2d} - f(z)z_h^{2d}}},
\end{equation}
as the reference time as shown in Fig.\ref{HM-RT}.

Plugging eq.(\ref{dvdz}) and eq.(\ref{Cv}) into eq.(\ref{SHMdef}), we finally obtain the entanglement entropy corresponding to the HM surface,
\begin{equation} \label{SHM}
  \mathcal{S}_{\rm HM} = \frac{l^d_{AdS} V_{d-1}}{4 G_N^{(d+2)}} \int_0^{z_M} \frac{z_M^d dz}{ z^d \sqrt{f(z) z_M^{2d} - f(z_M) z^{2d}} }.
\end{equation}

\subsection{Boundary RT Surface} \label{Boundary RT Surface}

\begin{figure*}[t]
  \centering
  \includegraphics[scale=0.6]{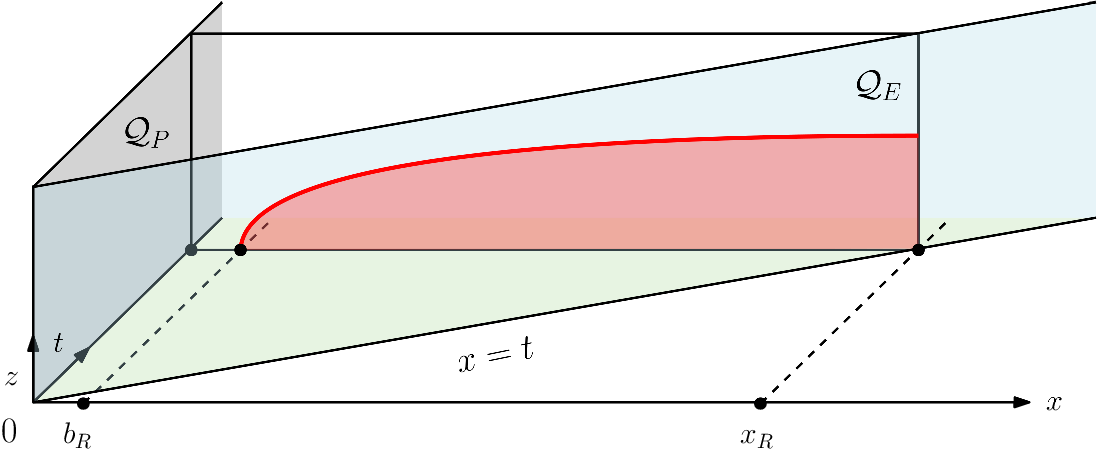}
  \caption{The red curve is the boundary RT surface ending on the ETW brane $\mathcal{Q}_E$ on a constant time slice in the static bulk spacetime.}
  \label{QPQE_2}
\end{figure*}

It was argued that for a time-dependent background the Hubeny-Rangamanni-Takayanagi (HRT) prescription must be used instead of the RT prescription so that the minimal surface can extend into the time direction \cite{0705.0016}. In our setup, the bulk spacetime is static, although the ETW brane is time-dependent. The time-dependence in our setup enters only through the location of where the HRT surface ends on the ETW brane. Since the bulk spacetime is static and enjoys the time reversal symmetry, we expect that the minimal surface obtained by using the HRT prescription is identical to the RT surface in a constant time slice as shown on Fig.\ref{QPQE_2}. In the following, we will show that this is true by the concrete calculation of the boundary effect. Some works utilizing a time-dependent brane in the braneworld model with the similar conclusion can be seen in \cite{2004.08020,2105.12614}.

The effective Hawking radiation region at a fixed time is,\footnote{Due to the symmetry between the left and right sides, we will only consider the right side as the example in the following calculations. The left side is exactly the same. The complete description of the entanglement entropy is the sum of that from both sides.}
\begin{equation} 
  \begin{split}
  \mathcal{R} = \left\{ x \in (b_R, x_R), \textbf{x}_{d-1} \in \mathbb{R}^{d-1} \right\},
  \end{split}
\end{equation}
which preserves $(d-1)$-dimensional translation invariance. Therefore, we can describe the BRT surface by $t=t(x)$ and $z = z(x)$, that leads to the following induced metric of the BRT surface,
\begin{equation}
ds^{2}=\frac{l_{AdS}^{2}}{z^{2}}\left[  \left( -f\left(  z\right)
t^{\prime2}+\frac{z^{\prime2}}{f\left(  z\right)  } + 1\right)  dx^{2}+\sum
_{i=1}^{d-1}dx_{i}^{2}\right],
\end{equation}
where the prime in this subsection denotes the derivative with respect to $x$.

Then the string action eq.(\ref{action}) becomes,

\begin{equation}
\mathcal{S}_{\rm BRT} = \frac{l_{AdS}^d V_{d-1}}{4 G_N^{(d+2)}} \int_{0}^{x_{\rm{R}}}dx  \mathcal{L}(t', z, z'), \label{SBRT-action}
\end{equation}
where $x_R$ is the location in $x$ direction where the BRT surface ends on the ETW brane, and the Lagrangian is

\begin{equation}
    \mathcal{L}(t', z, z') = \frac{1}{z^d}
\sqrt{-f(z) t^{\prime2} + \frac{z^{\prime2}}{f(z)} + 1} .\label{Lagrangian}
\end{equation}
For convenience, we define $t_R=t(x_R)$ and $z_R=z(x_R)$ in the following.

To study the boundary effect, we first reparameterize the BRT surface as $(t(\tau), z(\tau), x(\tau))$, with $0 < \tau <1$ as the parameter, where 0 indicates the point on the entangling subregion while 1 indicates the point on the ETW brane. The action then reads, 

\begin{equation}
\mathcal{S}_{\rm BRT} = \frac{l_{AdS}^d V_{d-1}}{4 G_N^{(d+2)}} \int_{0}^{1} \frac{d\tau}{z^d} \mathcal{L}(\dot{t},z,\dot{z},\dot{x}), \label{SBRT-action-affine}
\end{equation}
where the Lagrangian is

\begin{equation}
    \mathcal{L}(\dot{t},z,\dot{z},\dot{x}) = \frac{1}{z^d} \sqrt{-f(z) \dot{t}^2 + \frac{\dot{z}^2}{f(z)} + \dot{x}^2},
\end{equation}
and the dot denotes the derivative with  respect to $\tau$.
Varying the action eq.\eqref{SBRT-action-affine} we find (we drop the prefactor of the integral in the mean time to declutter the expression),

\begin{align}
    0 & = \delta S_{\rm{BRT}} \nonumber\\
    & = \int_0^1 \frac{d\tau}{z^d} \frac{\left( -f(z) \dot{t} \delta \dot{t} + \frac{\dot{z}}{f(z)} \delta \dot{z} + \dot{x} \delta \dot{x} \right)}{\sqrt{-f(z) \dot{t}^2 + \frac{\dot{z}^2}{f(z)} + \dot{x}^2}} + \int_0^1 \frac{\partial \mathcal{L}}{\partial z} \delta z d\tau \nonumber\\
    & = \frac{-f(z) \dot{t} \delta t + \frac{\dot{z}}{f(z)} \delta z + \dot{x} \delta x}{z^d \sqrt{-f(z) \dot{t}^2 + \frac{\dot{z}^2}{f(z)} + \dot{x}^2}} \Biggr|_0^1 - \int_0^1 \Sigma_i (\rm{EOM})_i \delta r_i d\tau,
\end{align}
where $r_i = \{t, z , x\}$ and $(\rm{EOM})_i$ are the corresponding Euler-Lagrange equations. The boundary condition imposed on the entangling subregion indicated by $\tau=0$ follows the standard procedure, e.g. Dirichlet condition, and we will not repeat it here. The important boundary condition that we will consider is the one on the ETW brane. In order for the boundary variation to vanish, we must have

\begin{align}
    & -f(z) \dot{t} \delta t + \frac{\dot{z}}{f(z)} \delta z + \dot{x} \delta x \Bigr|^{\rm{brane}} = 0.
\end{align}
The extremization in the context of the RT proposal for a static boundary leads to the boundary condition $\dot{z} = 0$ on the ETW brane. In our setup, the boundary, i.e. the ETW brane, is time dependent. Nevertheless, this would not change the boundary condition in the $z$ direction. So we have,

\begin{equation}
    -f(z) \dot{t} \delta t + \dot{x} \delta x \Bigr|^{\rm{brane}} = 0.
\end{equation}
The embedding equation of the ETW brane that the BRT surface ends on is given by $t=x$, then we have,

\begin{equation}
    \delta t = \delta x,
\end{equation}
which leads to the boundary condition,

\begin{equation}
    t' \Bigr|^{\rm{brane}} = \frac{dt}{dx} \Bigr|^{\rm{brane}} = \frac{\dot{t}}{\dot{x}} \Bigr|^{\rm{brane}} = \frac{1}{f(z_R)}. \label{bc1}
\end{equation}
As will be seen later, imposing this boundary condition will lead to the relevant quantities to not be defined. So in order to assure that the boundary variation vanishes, an alternative acceptable solution is to fix the point on the ETW brane along the t-x direction which means,

\begin{equation}
    \delta t = \delta x = 0. \label{bc2}
\end{equation}
Now, going back to eq.\eqref{SBRT-action}, we can solve the equation of motion and impose the corresponding boundary conditions.\footnote{Note that we can always switch from the dot to the primed system through the chain rule.}

The conjugate momenta are defined as,
\begin{eqnarray}
P_{t}& =& \frac{\partial\mathcal{L}}{\partial t^{\prime}}=\frac{-f\left(
z\right)  t^{\prime}}{z^{d}\sqrt{1-f\left(  z\right)  t^{\prime2}
+\frac{z^{\prime2}}{f\left(  z\right)  }}},\\
P_{z} &=& \frac{\partial\mathcal{L}}{\partial z^{\prime}}=\frac{z^{\prime}
}{z^{d}f\left(  z\right)  \sqrt{1-f\left(  z\right)  t^{\prime2}
+\frac{z^{\prime2}}{f\left(  z\right)  }}}.
\end{eqnarray}

The Hamiltonian then can be obtained as,
\begin{equation}
H =t'P_t+z'P_z-\mathcal{L}= \frac{-1}{z^d \sqrt{1 - f(z) t'^2 + \frac{z'^2}{f(z)}}}.
\end{equation}

Since the Lagrangian eq.(\ref{Lagrangian}) does not explicitly depend on the function $t(x)$,  $P_t$ and $H$ are conserved in this system, and implies that,
\begin{equation}
\frac{-f\left(  z\right)  t^{\prime}}{z^{d}\sqrt{1-f\left(  z\right)
t^{\prime2}+\frac{z^{\prime2}}{f\left(  z\right)  }}} = \frac{-f\left(
z_R\right)  t_{R}^{\prime}}{z_R^{d}\sqrt{1-f\left(  z_R\right)  t_{R}^{\prime2}+\frac{z_R^{\prime2}}{f\left(  z_R\right)}}},
\end{equation}

\begin{equation}
\frac{-1}{z^{d}\sqrt{1-f\left( z\right)  t^{\prime2}+\frac{z^{\prime2}
}{f\left(  z\right)}}} = \frac{-1}{z_R^{d}\sqrt{1-f\left(z_R\right)  t_{R}^{\prime2}+\frac{z_R^{\prime2}}{f\left( z_R\right)}}}.
\end{equation}
As can be seen, after imposing $z'_R = 0$ the conserved quantities are only well defined for $1-f(z_R) t'^2_R >0$ which rules out the choice of the boundary condition eq.(\ref{bc1}). Thus we have to use the other boundary condition eq.(\ref{bc2}) of fixing the point along the t-x direction on the ETW brane that leads to an allowed solution of the variational problem.

\begin{figure}
  \centering
  \includegraphics[scale=0.4]{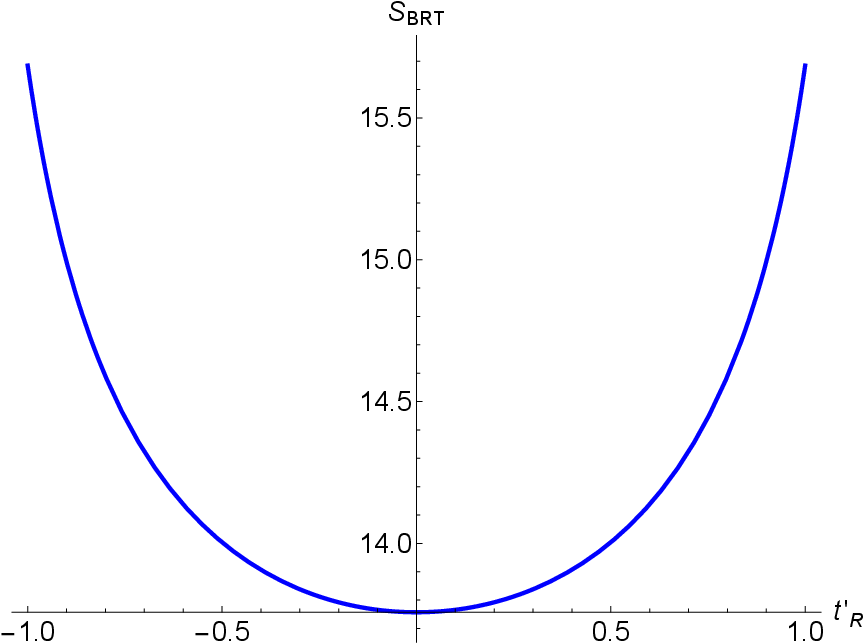}
  \caption{Plot of the extremal surface as a function of $t'_{R}$.}
  \label{SBRT-tp plot}
\end{figure}

The corresponding $t'$ and $z'$ can be written in terms of the quantities on the brane,
\begin{widetext}
\begin{equation}
t^{\prime} = \frac{f\left(  z_{\rm{R}}\right)  }{f\left(  z\right)  }t_{R}^{\prime}, \quad
z^{\prime} = \sqrt{f\left(  z\right)  \left(  \frac{z_R^{2d}}{z^{2d}
}-1\right)  -f\left(  z_R\right)^{2}  \left[  \frac{z_R^{2d}f\left(
z\right)  }{z^{2d}f\left(z_R\right)}-1\right]  t_{R}^{\prime
2}+\frac{z_R^{2d}f\left(  z\right)  }{z^{2d}f\left(  z_R\right)
}z_R^{\prime2}}.
\label{tp}
\end{equation}
\end{widetext}
The extremal surface is then obtained to be,
\begin{equation}
\mathcal{S}_{\rm BRT} = \frac{l_{AdS}^d V_{d-1}}{4 G_N^{(d+2)}} \int_{0}^{z_R}\frac{dz}{z^{d}\sqrt
{f(z) \left(1-\frac{z^{2d}}{z_R^{2d}}\frac{1-\frac{f\left(  z_R\right)^2
}{f\left(  z\right)  }t_{R}^{\prime2}}{1-f\left(  z_R\right)  t_{R}^{\prime2}}\right)}}. \label{SBRT-tp}
\end{equation}
According to the HRT prescription, given a family of extremal surfaces, we must pick the one with the minimal value. The extremal surface has a minimal value at $t'_{R} = 0$ as can be seen in Fig.\ref{SBRT-tp plot}, this leads to $t'(x)=0$ by eq.(\ref{tp}). This confirms that the BRT surface is on the constant-time slice, thus the HRT surface is identical to the RT surface. This is indeed the case since the conjugate momentum $P_t$ is conserved. Also, since the ETW brane has no backreaction, it cannot induce any time dependence in the bulk. For a time dependent boundary, e.g. the ETW brane, the HRT surface is not necessary to be orthogonal to the boundary as it should be for a static boundary. The similar conclusion has been obtained in \cite{2004.08020,2105.12614}.

The corresponding holographic entanglement entropy is given by,

\begin{equation} \label{SBRT}
  \mathcal{S}_{\rm BRT} = \frac{l_{AdS}^d V_{d-1}}{4 G_N^{(d+2)}} \int_0^{z_R} \frac{z_R^{d} dz}{z^d \sqrt{f(z) (z_R^{2d} - z^{2d})}},
\end{equation}
By integrating $z'$ in eq.\eqref{tp}, the length of the effective radiation region can be calculated,
\begin{equation} \label{azB}
  x_R - b_R = \int_0^{z_R} \frac{z^d dz}{\sqrt{f(z) (z_R^{2d} - z^{2d})}}.
\end{equation}
The entanglement entropy defined in eq.(\ref{SBRT}) is divergent near $z=0$ and so for the purposes of our numerical calculations we regularize it as,
\begin{equation} \label{SBRT-reg}
  \begin{split}
  \mathcal{S}_{\rm BRT-reg} = & \frac{l_{AdS}^d V_{d-1}}{4 G_N^{(d+2)}} \Bigg[ \int_0^{z_R} \frac{z_R^{d} dz}{z^d \sqrt{f(z) (z_R^{2d} - z^{2d})}} \\
  & - \int_0^{z_R} \frac{z_R^{d} dz}{z^d \sqrt{(z_R^{2d} - z^{2d})}} \Bigg].
  \end{split}
\end{equation}
We note that the entanglement entropies corresponding to the HM surface which we calculated in the last subsection, and the IRT surface which we will calculate in the next subsection, have the same divergence structure. Since we are only concerned with the difference between the entanglement entropies, the divergences will cancel each other out so that we do not need to regularize the other ones. The regularization of the BRT in eq.(\ref{SBRT-reg}) is only for the purpose of numerical plotting.


\begin{figure*}
	\subfloat[BRT-IRT]{\includegraphics[scale=0.27]{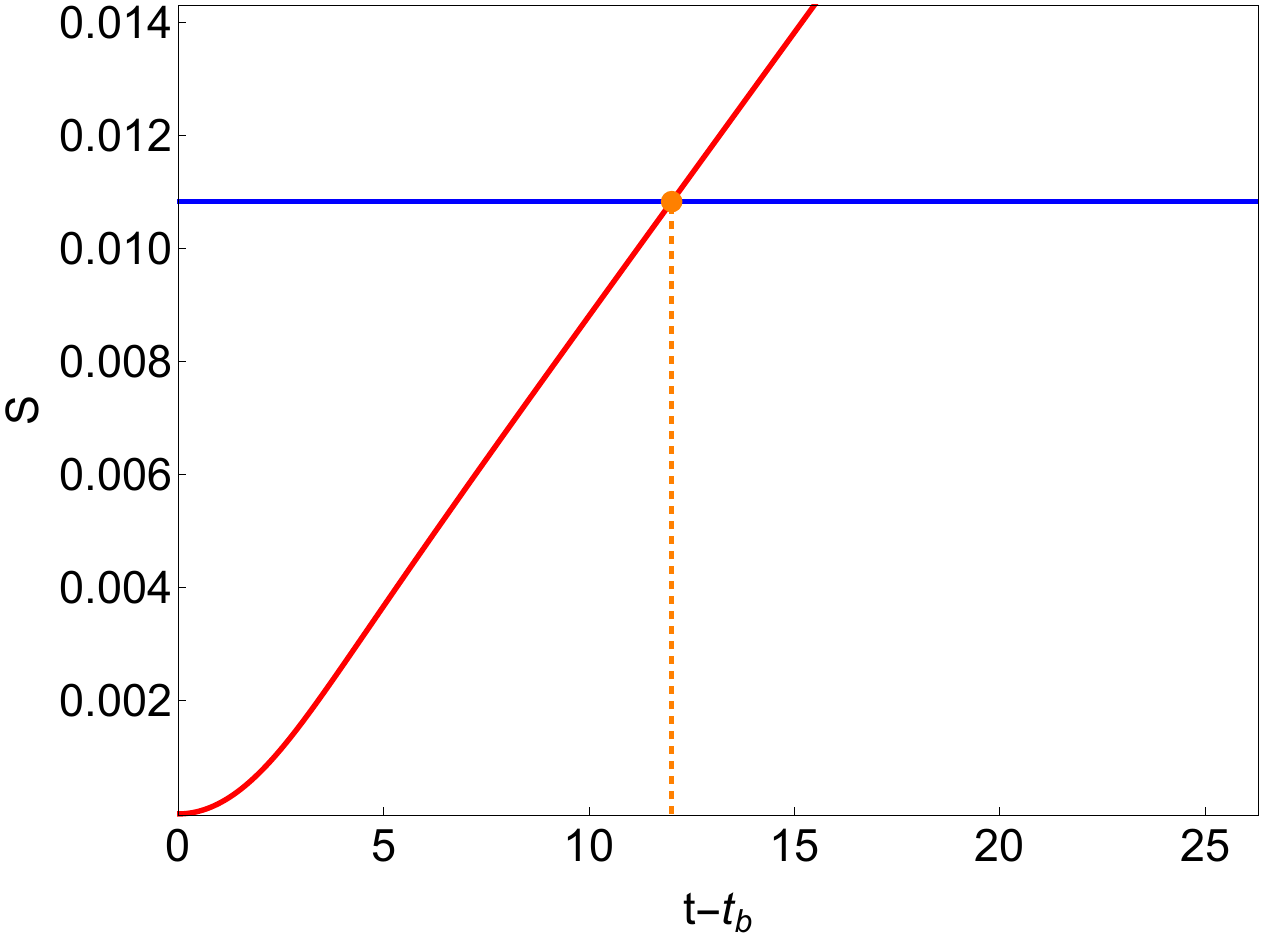}}
	\hfill
	\subfloat[HM-IRT]{\includegraphics[scale=0.27]{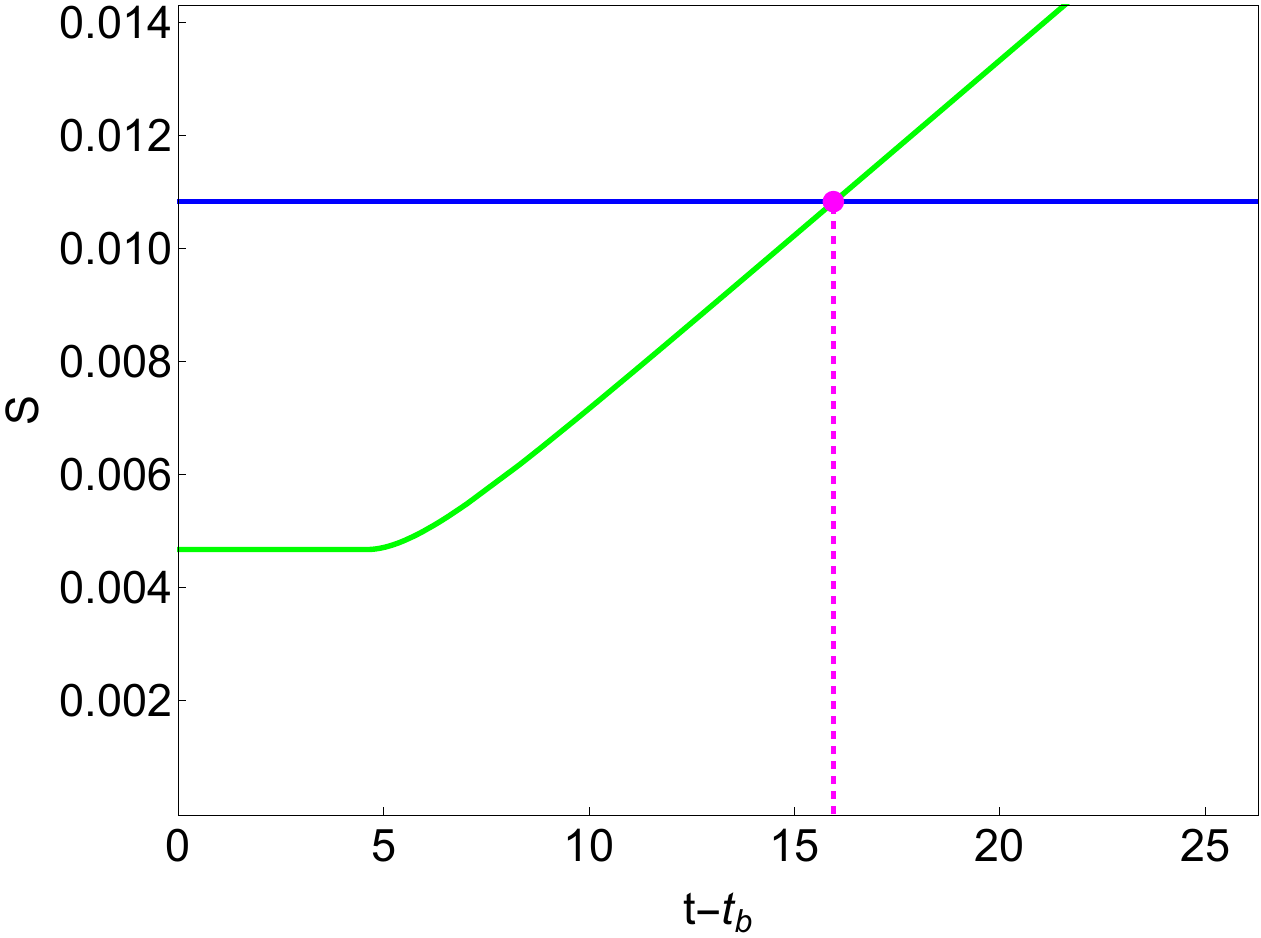}}
	\hfill
	\subfloat[BRT-HM]{\includegraphics[scale=0.27]{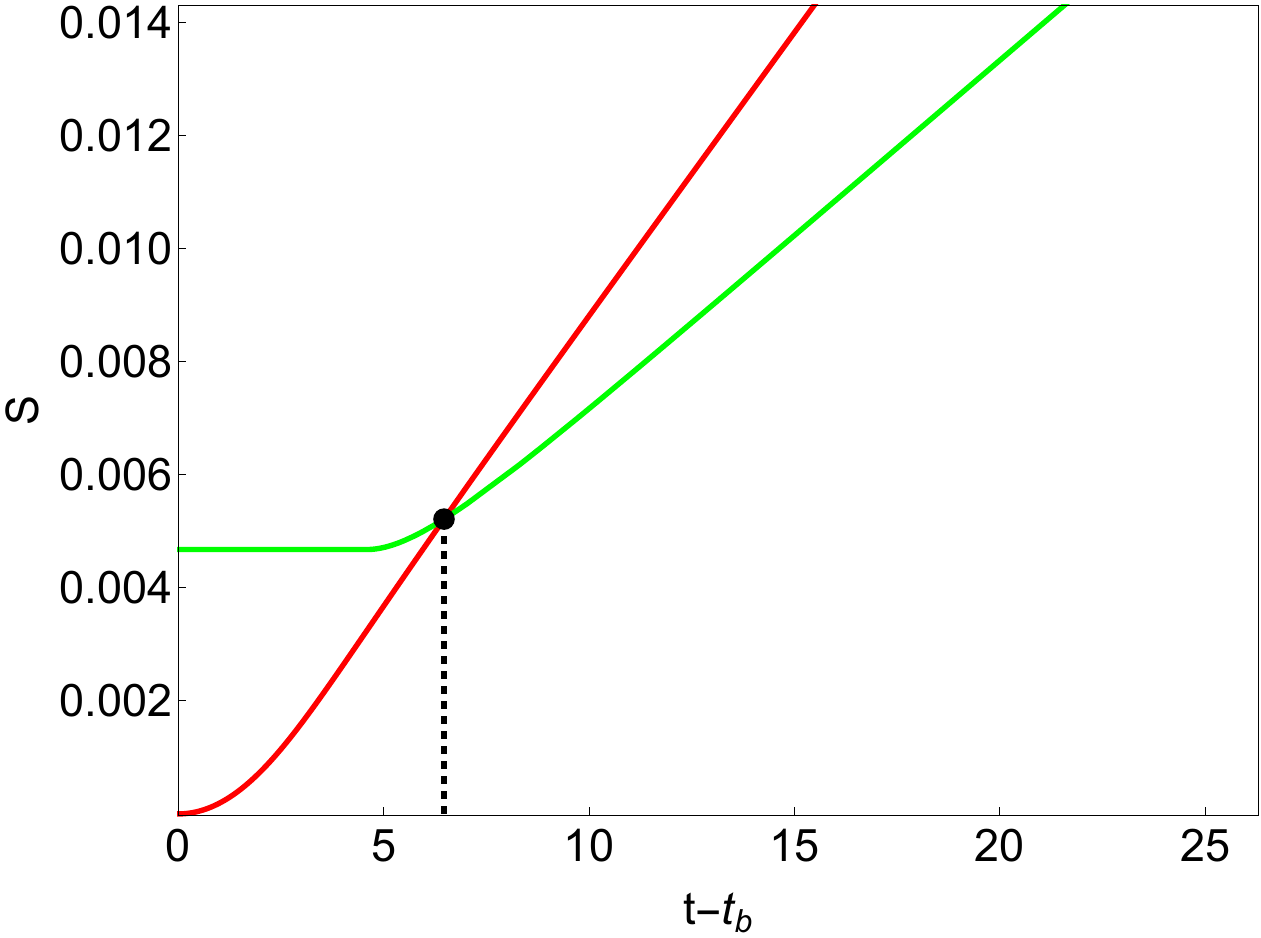}}
	\caption{The phase transitions between BRT (red), HM (green), and IRT (blue) surfaces for $z_h=10$. (a) $\mathcal{S}_{\rm BRT}$ and $\mathcal{S}_{\rm IRT}$. (b) $\mathcal{S}_{\rm HM}$ and $\mathcal{S}_{\rm IRT}$. (c) $\mathcal{S}_{\rm BRT}$ and $\mathcal{S}_{\rm HM}$.}
	\label{Phase_Diagrams_zh10}
\end{figure*}

\subsection{Island RT Surface} \label{HEE at Late Time and the Island}

The entanglement entropy of the IRT surface on the $z-x$ plane reads,
\begin{equation} \label{SIRTdef}
  \mathcal{S}_{\rm IRT} = \frac{l^d_{AdS} V_{d-1}}{4 G_N^{(d+2)}}\left[ \int dx \frac{1}{z^d} \sqrt{ 1 + \frac{1}{f(z)} \cdot \left( \frac{dz}{dx} \right)^2}+ \frac{1}{z_I^{d-1}} \right],
\end{equation}
which is similar to $\mathcal{S}_{\rm BRT}$ in eq.(\ref{SBRT-action}) with an extra term representing the contribution from the QES. Also, $z_{\rm QES} = z_I$ is the location of the QES on the Planck brane as shown in Fig.\ref{Penrose 1}.
In a similar way to section \ref{Boundary RT Surface}, we can define a constant $C_\sigma$,
\begin{equation} \label{Csig}
  C_\sigma = \frac{-1}{z^d \sqrt{1 + \frac{z'^2}{f(z)}}},
\end{equation}
and solve for $z'$ as,
\begin{equation} \label{dzdxIRT}
  z' = \pm \frac{\sqrt{f(z) \left( \frac{1}{C_\sigma^2} - z^{2d} \right)}}{z^d}.
\end{equation}
However, in the presence of the QES, the IRT surface does not need to be perpendicular to the Planck brane. The boundary conditions now are given by,
\begin{equation} \label{BC of IRT}
  z(x = 0) = z_I, \quad z(x = b_R) = 0,
\end{equation}
With the above boundary conditions, we get,
\begin{equation}
  \frac{dz}{dx} = \pm \frac{\sqrt{f(z) \left( z_I^{2d} - z^{2d} + \frac{z_I^{2d} \sigma^2}{f(z_I)} \right)}}{z^d},
\end{equation}
which can be integrated to give,
\begin{equation} \label{bzI}
  b_R = \int_0^{z_I} \frac{z^d dz}{\sqrt{f(z) \left( z_I^{2d} - z^{2d} + \frac{z_I^{2d} \sigma^2}{f(z_I)} \right)}},
\end{equation}
where $\left. \sigma=dz/dx \right|_{x = 0} $ is the slope as the IRT surface reaches the Planck brane at $x=0$. The slope $\sigma$ can be determined by finding the location $z_I$ of the QES on the Planck brane that minimizes $\mathcal{S}_{\rm IRT}$ for a fixed cutoff $b_{R}$. Without the QES contribution, the minimum condition implies $\sigma=0$, i.e. the RT surface is perpendicular to the brane as we have known.

Finally, the entanglement entropy corresponding to the IRT surface eq.\eqref{SIRTdef} can be expressed as,
\begin{equation} \label{SIRT}
  \begin{split}
  & \mathcal{S}_{\rm IRT} = \frac{l^d_{AdS} V_{d-1}}{4 G_N^{(d+2)}} \\
  & \left[ \int_0^{z_I} \frac{z_I^d}{z^d} \sqrt{\frac{\left( 1 + \frac{\sigma^2}{f(z_I)} \right)}{f(z) \left( z_I^{2d} - z^{2d} + \frac{z_I^{2d} \sigma^2}{f(z_I)} \right)}} dz + \frac{1}{z_I^{d-1}} \right].
  \end{split}
\end{equation}
Now, the condition for the location $z_I$ of the QES on the Planck brane that minimizes $\mathcal{S}_{\rm IRT}$ is,
\begin{equation} \label{dS/dz}
    \frac{d \mathcal{S}_{\rm IRT} }{d z_I}=0.
\end{equation}
Combining eq.\eqref{bzI} and eq.\eqref{dS/dz}, we can solve for $z_I$ as well as $\sigma$ for a fixed $b_R$.


\section{Phase Transitions and Page Curve} \label{Phase Transition and Page Curves}
As we have seen in the last section, there are three RT surfaces in the eternal black hole system, i.e. the HM, BRT, and IRT surfaces, whose corresponding entanglement entropies have been calculated in eq.\eqref{SHM}, eq.\eqref{SBRT}, and eq.\eqref{SIRT}. The dominant one is the minimum among them,

\begin{equation}
  \mathcal{S}=\min(\mathcal{S}_{\rm HM},\mathcal{S}_{\rm BRT},\mathcal{S}_{\rm IRT}).
\end{equation}
Among the three entanglement entropies, $\mathcal{S}_{\rm IRT}$ is a constant at a fixed temperature, while the other two increase with time. At the initial time $t=t_b$, $\mathcal{S}_{\rm BRT}=0$ dominates. At a later time, there could be phase transitions between each pair of the three entanglement entropies. In this section, we will investigate the phase transitions in detail and obtain the Page curve for the entanglement entropy of the eternal black hole.

\subsection{Phase Transitions} \label{Phase Transitions}

\begin{figure*}[t]
    \centering
	\subfloat[$\rm z_h = 10$]{\includegraphics[height=4.5cm]{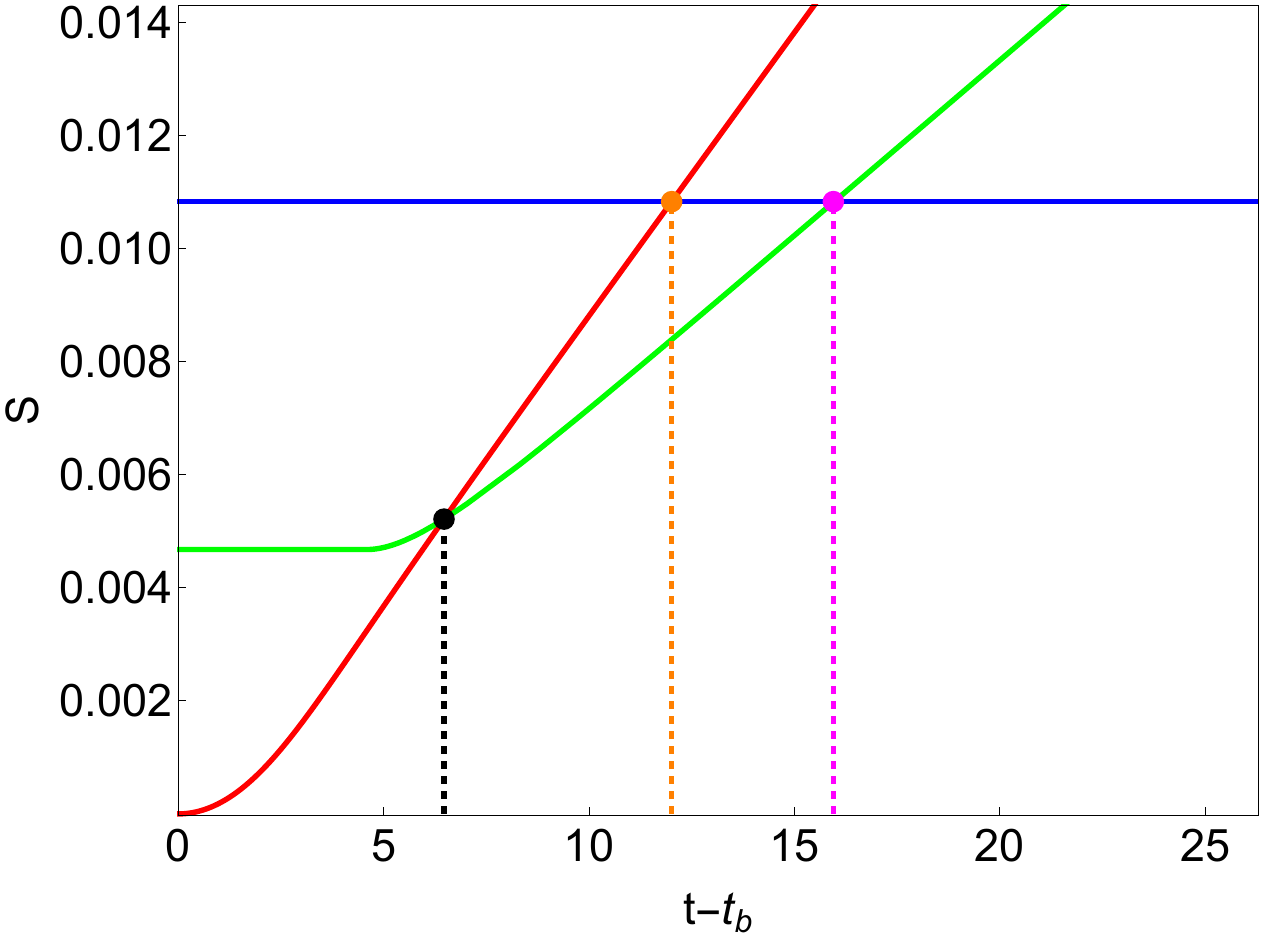}}
	\label{Phase_Transition_zh10}
	\hspace{1cm}
	\subfloat[$\rm z_h = 0.5$]{\includegraphics[height=4.5cm]{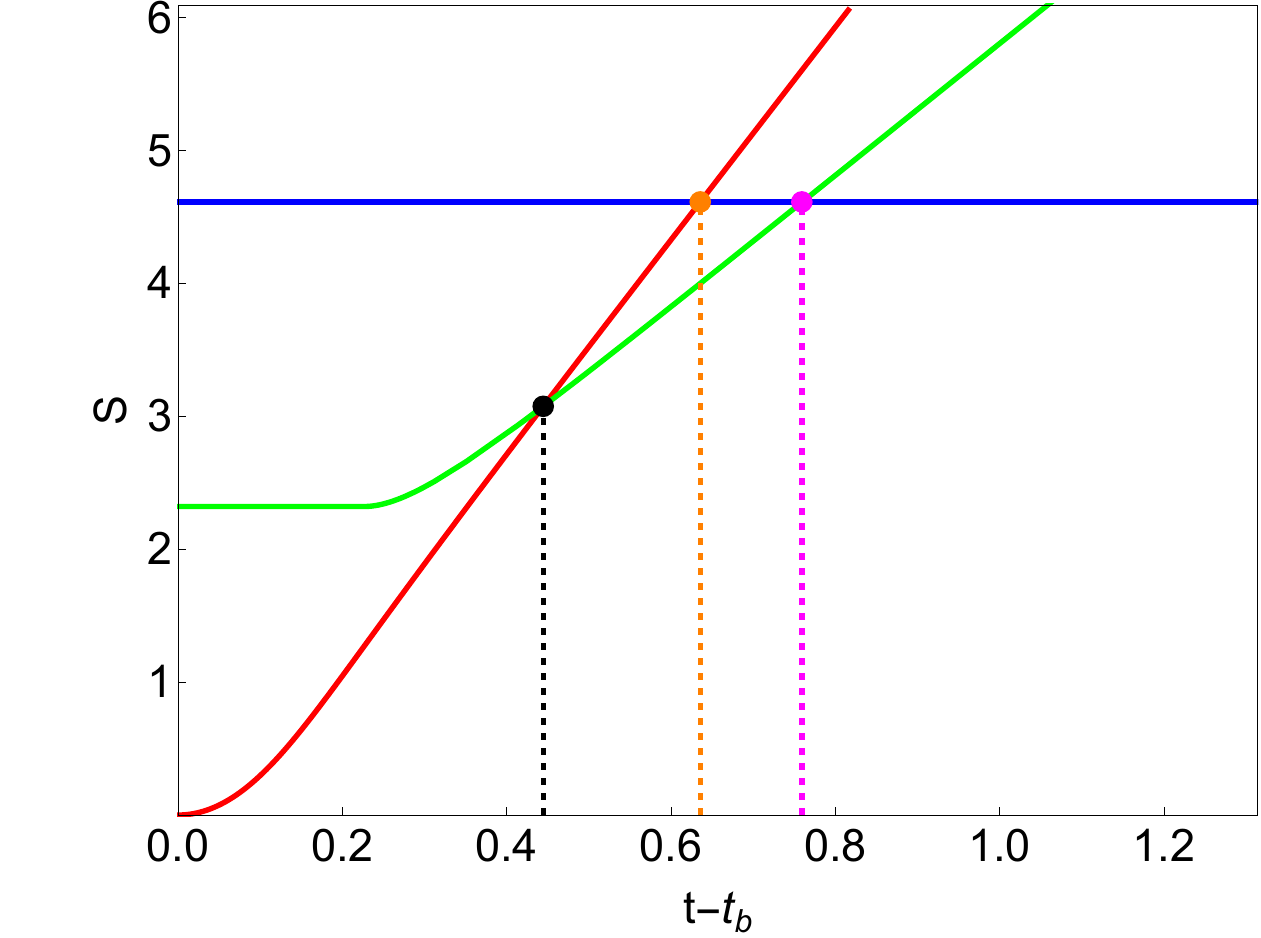}}
    \label{Phase_Transition_zh05}
    \subfloat[$\rm z_h = 0.1355$]{\includegraphics[height=4.5cm]{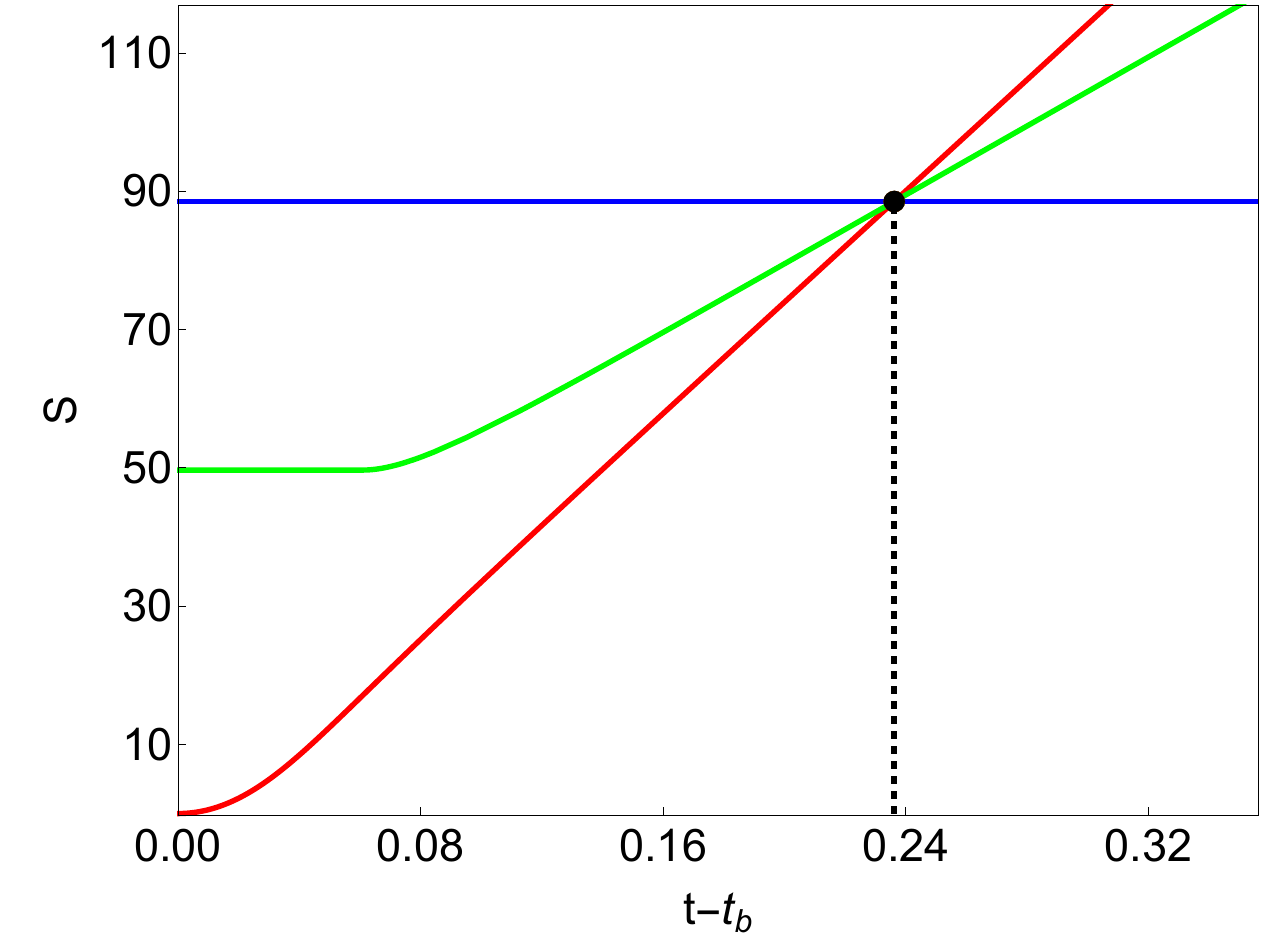}}
    \label{Phase_Transition_zh01355}
    \hspace{1cm}
    \subfloat[$\rm z_h = 0.1$]{\includegraphics[height=4.5cm]{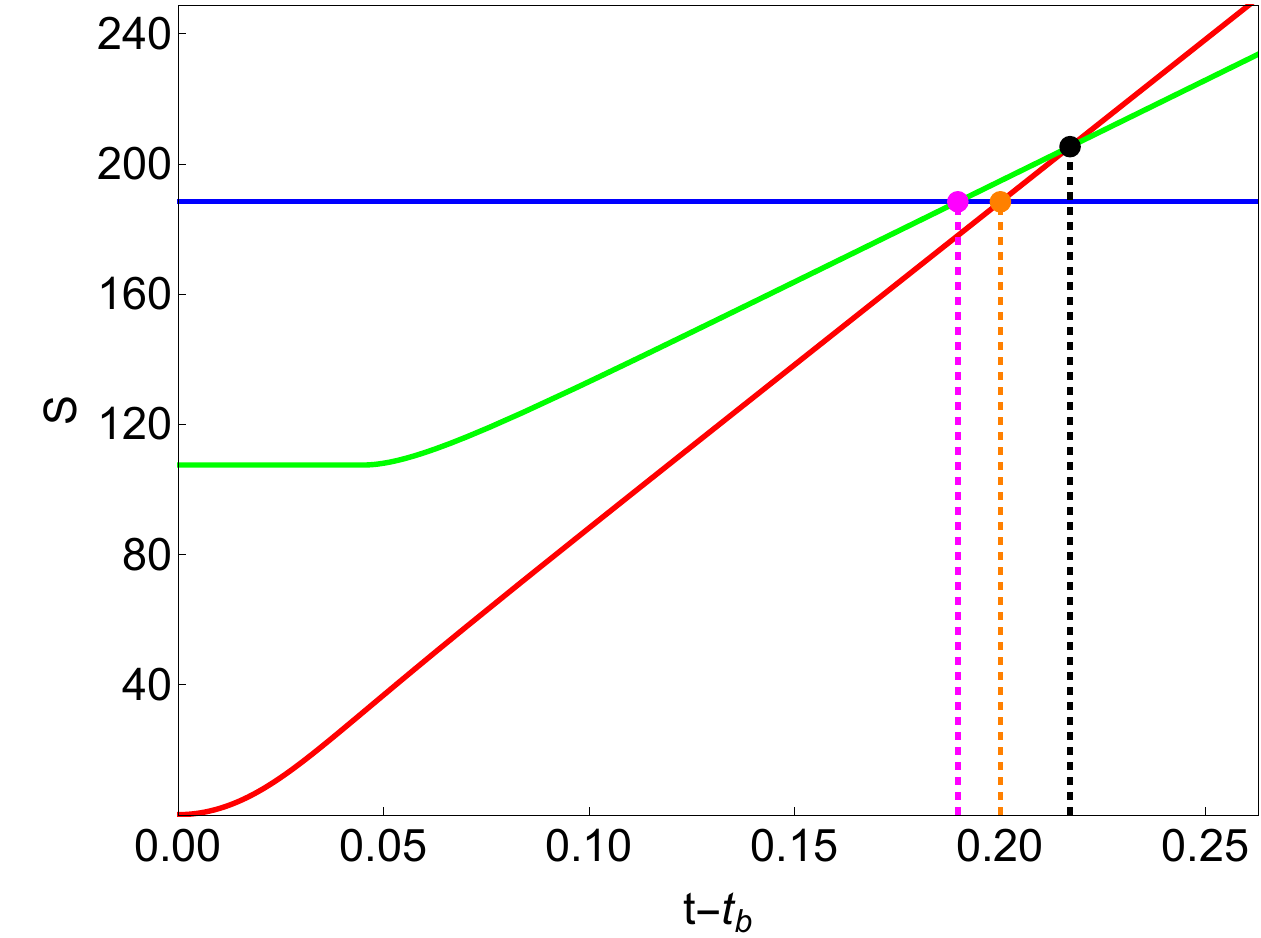}}
    \label{Phase_Transition_zh01}
    \caption{The behavior of the entanglement entropies, $\mathcal{S}_{\rm BRT}$ (red), $\mathcal{S}_{\rm HM}$ (green), and $\mathcal{S}_{\rm IRT}$ (blue), at different temperatures or black hole horizons. The black, orange, and violet dots indicate the phase transitions of $\mathcal{S}_{\rm BRT}\sim \mathcal{S}_{\rm HM}$, $\mathcal{S}_{\rm BRT}\sim \mathcal{S}_{\rm IRT}$, and $\mathcal{S}_{\rm HM}\sim \mathcal{S}_{\rm IRT}$, respectively.}
    \label{HEE at different T}
\end{figure*}

To be concrete in this subsection, we set the cutoff $b_L=b_R=b=0.1$, and the black hole horizon $z_h =10$ which corresponds to the temperature $T=0.032$.

Since the entanglement entropy corresponding to the IRT surface is a time independent constant, we will first consider the phase transitions between the IRT surface and the other two surfaces.
The phase transition between $\mathcal{S}_{\rm BRT}$ and $\mathcal{S}_{\rm IRT}$ can be obtained from

\begin{equation} \label{SBRT-SIRT}
  \Delta \mathcal{S}_{\rm BRT-IRT} = \mathcal{S}_{\rm BRT} - \mathcal{S}_{\rm IRT} = 0,
\end{equation}
where $\mathcal{S}_{\rm BRT}$ and $\mathcal{S}_{\rm IRT}$ are given in eq.\eqref{SBRT} and eq.\eqref{SIRT}.

The regularized\footnote{We regularize the entanglement entropy using eq.\eqref{SBRT-reg} to plot it numerically. However, we should note that the exact phase transition time is determined from eq.\eqref{SBRT-SIRT} without regularization.} entanglement entropy corresponding to the BRT and IRT surfaces are plotted in Fig.\ref{Phase_Diagrams_zh10}(a).
$\mathcal{S}_{\rm BRT}$ (red line) starts from zero at $t=t_b$ and increases almost linearly. It dominates until $t-t_b=t_{\rm BRT-IRT} \simeq 12.01$ when the phase transition between $\mathcal{S}_{\rm BRT}$ and $\mathcal{S}_{\rm IRT}$ (blue line) takes place, then $\mathcal{S}_{\rm IRT}$ will dominate for $t-t_b>t_{\rm BRT-IRT}$.

Similarly, the phase transition between $\mathcal{S}_{\rm HM}$ and $\mathcal{S}_{\rm IRT}$ can be obtained from

\begin{equation} \label{SHM-SIRT}
  \Delta \mathcal{S}_{\rm HM-IRT} = \mathcal{S}_{\rm HM} - \mathcal{S}_{\rm IRT} = 0,
\end{equation}
where $\mathcal{S}_{\rm HM}$ and $\mathcal{S}_{\rm IRT}$ are given in eq.\eqref{SHM} and eq.\eqref{SIRT}.

The regularized entanglement entropy corresponding to the HM and IRT surfaces are plotted in Fig.\ref{Phase_Diagrams_zh10}(b).
$\mathcal{S}_{\rm HM}$ (green line) starts from a finite value at $t=t_b$ and increases. It dominates until $t-t_b=t_{\rm HM-IRT}\simeq 15.95$ when the phase transition between $\mathcal{S}_{\rm HM} $ and $\mathcal{S}_{\rm IRT} $ takes place, then $\mathcal{S}_{\rm IRT}$ will dominate for $t-t_b>t_{\rm HM-IRT}$.

Finally, there is also a phase transition between $\mathcal{S}_{\rm BRT}$ and $\mathcal{S}_{\rm HM}$ that can be obtained from

\begin{equation} \label{SBRT-SHM}
  \Delta \mathcal{S}_{\rm BRT-HM} = \mathcal{S}_{\rm BRT} - \mathcal{S}_{\rm HM} = 0,
\end{equation}
where $\mathcal{S}_{\rm BRT}$ and $\mathcal{S}_{\rm HM}$ are given in eq.\eqref{SBRT} and eq.\eqref{SHM}.

The regularized entanglement entropy corresponding to the BRT and HM surfaces are plotted in Fig.\ref{Phase_Diagrams_zh10}(c). Both $\mathcal{S}_{\rm BRT}$ and
$\mathcal{S}_{\rm HM}$ increases with time. Among the two, $\mathcal{S}_{\rm BRT}$ dominates until $t-t_b=t_{\rm BRT-HM}\simeq 6.47$ when the phase transition between $\mathcal{S}_{\rm BRT}$ and $\mathcal{S}_{\rm HM}$ takes place, then $\mathcal{S}_{\rm HM}$ will dominate for $t-t_b>t_{\rm BRT-HM}$.

We have obtained the phase transitions of every pair of the three entanglement entropies corresponding to the BRT, HM, and IRT surfaces for the black hole horizon $z_h=10$ or $T=0.032$.
Now we put them together in Fig.\ref{HEE at different T}(a) with the red/green/blue line representing $\mathcal{S}_{\rm BRT}$/$\mathcal{S}_{\rm HM}$/$\mathcal{S}_{\rm IRT}$. The three critical points are labeled by black, orange, and violet dots.

At early times, $\mathcal{S}_{\rm BRT}$ dominates the system. Later on, at $t-t_b \simeq 6.47$, the phase transition between $\mathcal{S}_{\rm BRT}$ and $\mathcal{S}_{\rm HM}$ takes place, and $\mathcal{S}_{\rm HM}$ becomes dominant. Then, at $t-t_b \simeq 15.95$, the phase transition between $\mathcal{S}_{\rm HM}$ and $\mathcal{S}_{\rm IRT}$ occurs, and $\mathcal{S}_{\rm IRT}$ becomes dominant. In addition, there is another phase transition between $\mathcal{S}_{\rm BRT}$ and $\mathcal{S}_{\rm IRT}$ that occurs between the above two phase transitions at $t-t_b \simeq 12.01$. However, this phase transition is irrelevant since it will not occur in the real physical process as shown in Fig.\ref{HEE at different T}(a).

Having obtained the phase diagram of the entanglement entropy at $z_h=10$ or $T=0.032$, let us consider the phase diagrams at other temperatures. Fig.\ref{HEE at different T}(b) shows the entanglement entropies for the black hole horizon $z_h=0.5$, which corresponds to a higher temperature $T=0.637$. As in the case of $z_h=10$, there are three phase transitions in this case. The phase transitions of $\mathcal{S}_{\rm BRT}\sim \mathcal{S}_{\rm HM}$, $\mathcal{S}_{\rm BRT}\sim \mathcal{S}_{\rm IRT}$, and $\mathcal{S}_{\rm HM}\sim \mathcal{S}_{\rm IRT}$ takes place at $t-t_b \simeq 0.444$, $t-t_b \simeq 0.636$ and $t-t_b \simeq 0.759$, respectively.

\begin{figure*}[t]
    \centering
	\subfloat[$T<T_c$]{\includegraphics[width=0.3\textwidth]{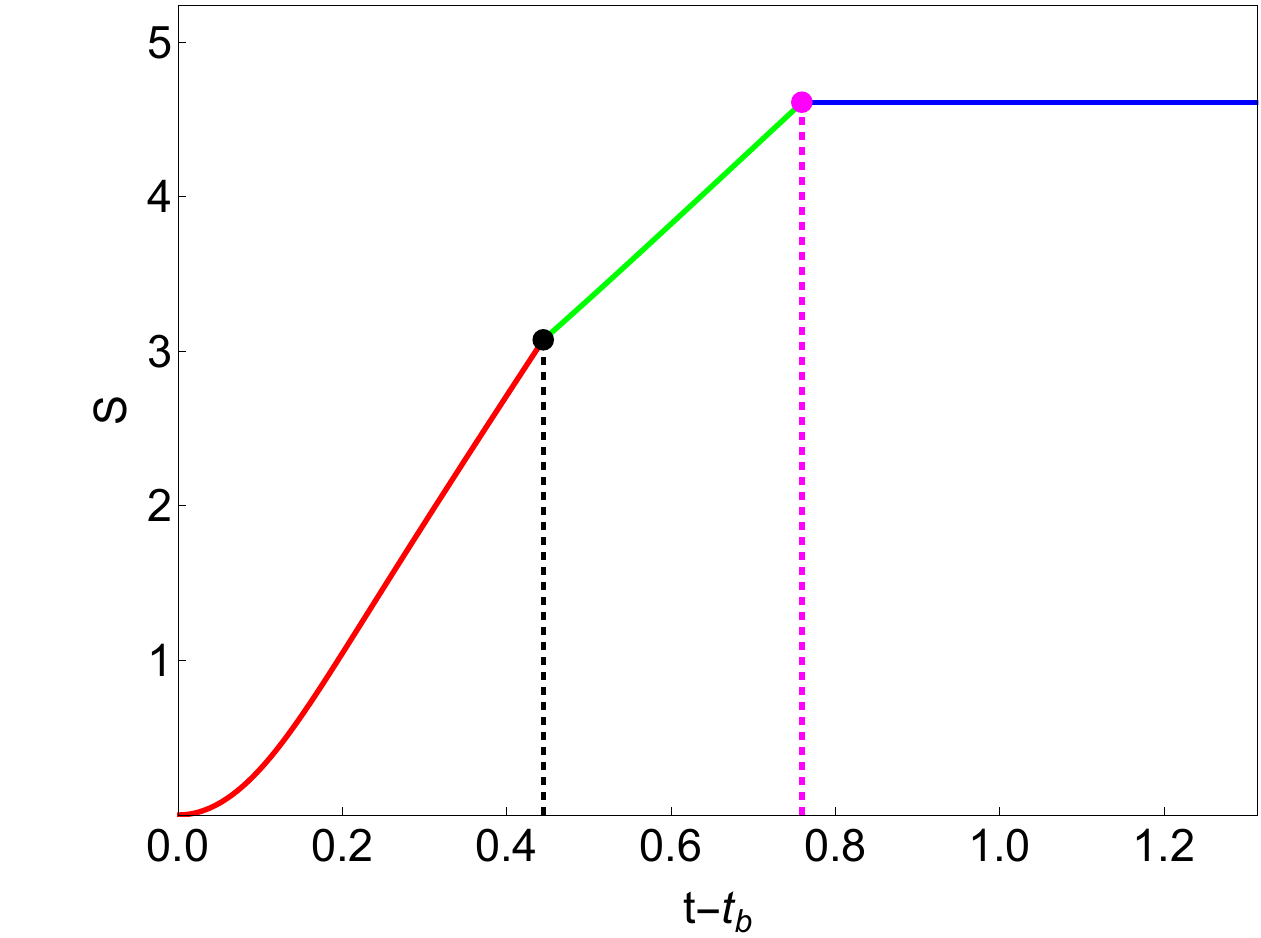}}
	\hfill
	\subfloat[$T=T_c$]{\includegraphics[width=0.3\textwidth]{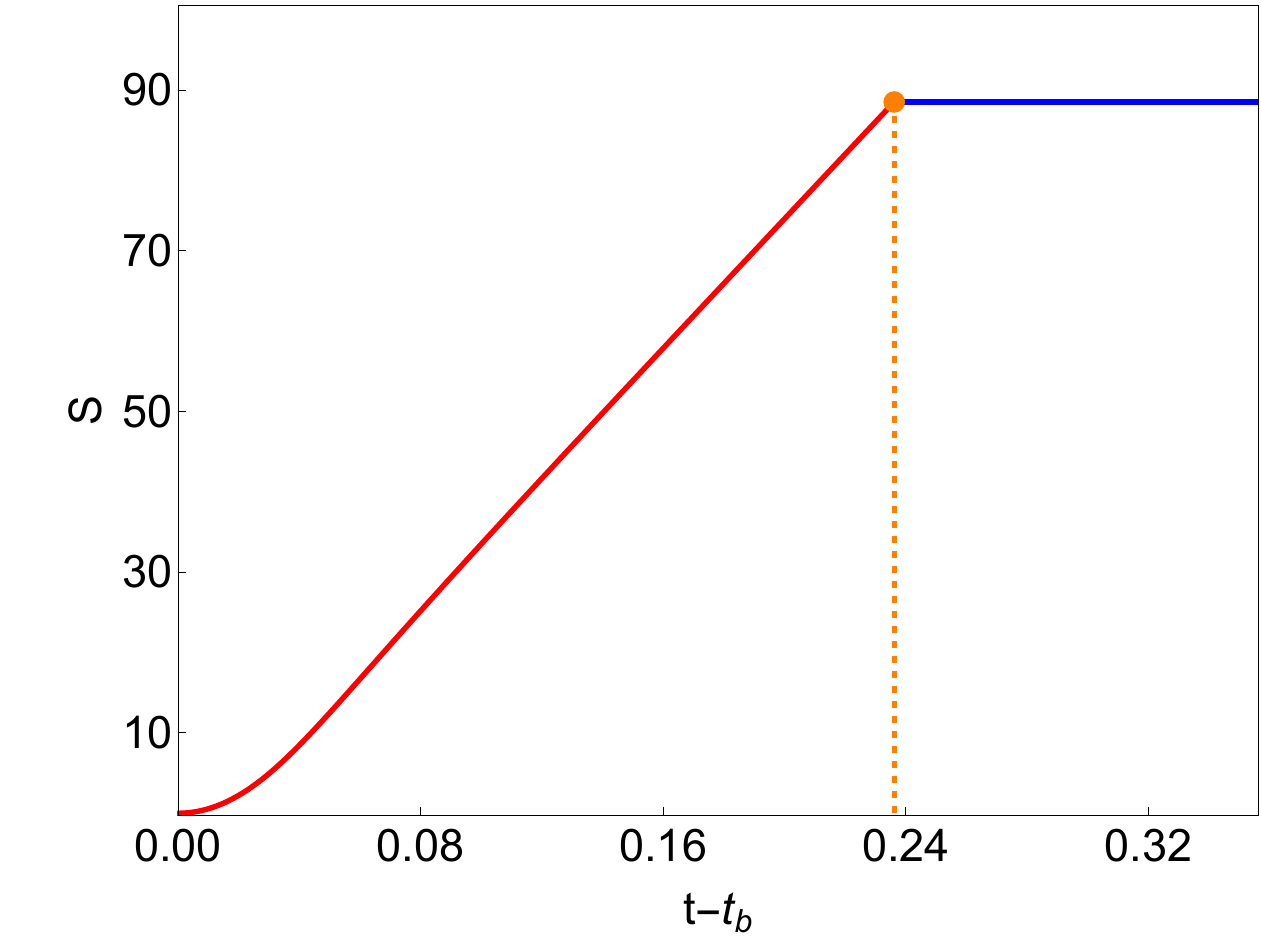}}
	\hfill
	\subfloat[$T>T_c$]{\includegraphics[width=0.3\textwidth]{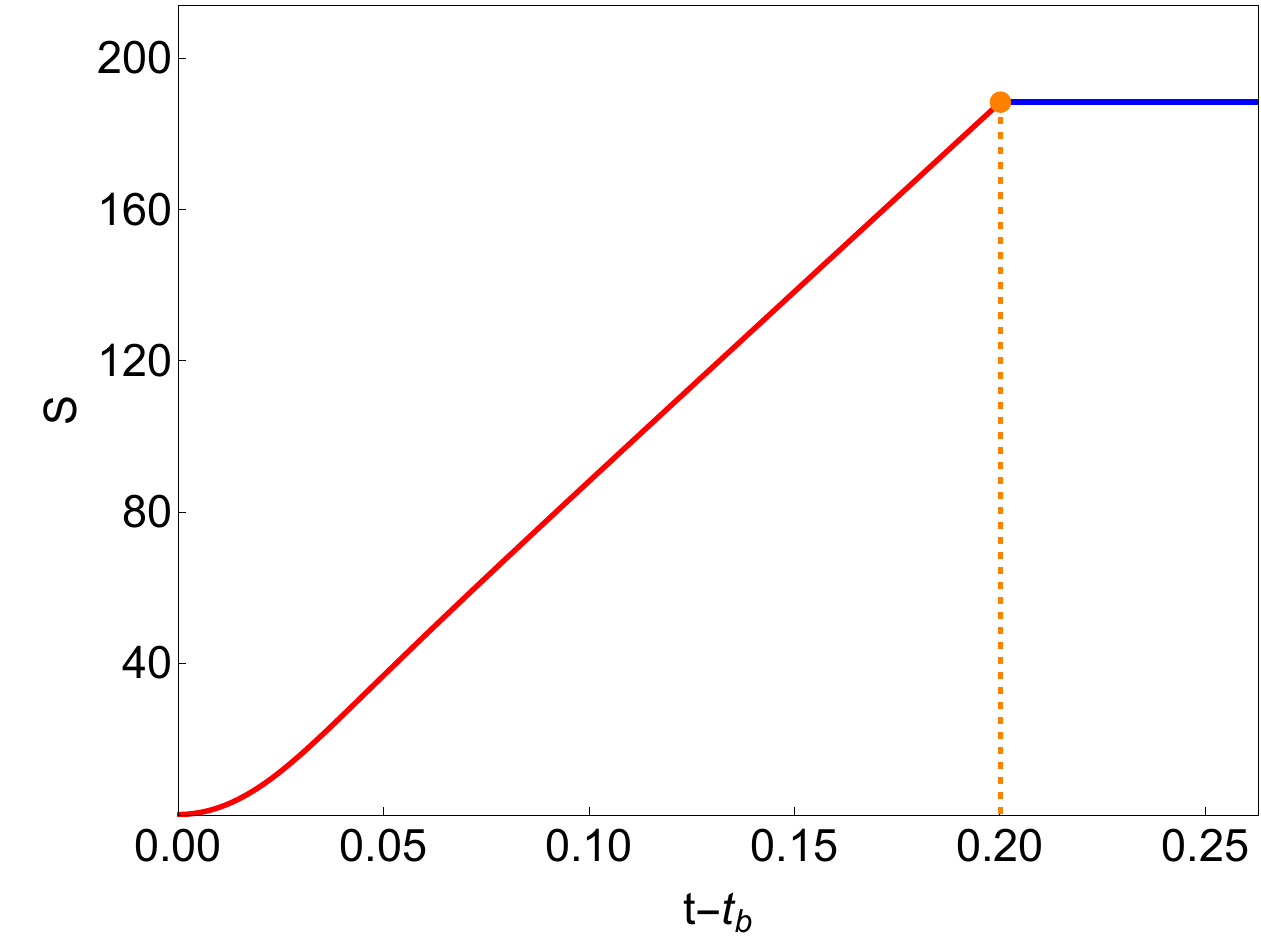}}
	\caption{The Page curve for different temperatures. (a) For the case of $z_h = 0.5$, $\mathcal{S}_{\rm BRT}$ transitions to $\mathcal{S}_{\rm HM}$ at $t-t_b \simeq 0.444$, then to $\mathcal{S}_{\rm IRT}$ at $t-t_b \simeq 0.759$. (b) For the case of $z_h = 0.1355$, $\mathcal{S}_{\rm BRT}$ transitions directly to $\mathcal{S}_{\rm IRT}$ at $t-t_b \simeq 0.236$. (c) For the case of $z_h = 0.1$, $\mathcal{S}_{\rm BRT}$ transitions directly to $\mathcal{S}_{\rm IRT}$ at $t-t_b \simeq 0.200$.}
	\label{Page_Curve}
\end{figure*}

The phase structure for $z_h=0.5$ is similar to that for $z_h=10$. Nevertheless, we have an important observation that the three critical points become closer for a smaller horizon or a higher temperature, so that the duration of $\mathcal{S}_{\rm HM}$ domination shrinks. Therefore, an interesting question is whether or not the duration of $\mathcal{S}_{\rm HM}$ would shrink to zero if the temperature continuously increases?  We found that the answer is yes.

Fig.\ref{HEE at different T}(c) shows the entanglement entropies for the black hole horizon $z_h=0.1355$, which corresponds to the critical temperature $T_c=2.349$. At this temperature, the three phase transitions coincide at the same point at $t-t_b \simeq 0.236$ when the time duration of $\mathcal{S}_{\rm HM}$ shrinks to zero, and $\mathcal{S}_{\rm BRT}$ transit to $\mathcal{S}_{\rm IRT}$ directly at this critical point.

For $T\ge T_c$, for example, in the case of the black hole horizon $z_h=0.1$, which corresponds to the temperature $T=3.183\ge T_c$, the entanglement entropies are shown in Fig.\ref{HEE at different T}(d). The phase transitions of $\mathcal{S}_{\rm BRT}\sim \mathcal{S}_{\rm HM}$, $\mathcal{S}_{\rm BRT}\sim \mathcal{S}_{\rm IRT}$, and $\mathcal{S}_{\rm HM}\sim \mathcal{S}_{\rm IRT}$ take place at $t-t_b \simeq 0.217$, $t-t_b \simeq 0.200$ and $t-t_b \simeq 0.190$, respectively. We notice that the order of the three phase transitions reverses. Because the phase transition between $\mathcal{S}_{\rm BRT}$ and $\mathcal{S}_{\rm HM}$ occurs after that between $\mathcal{S}_{\rm BRT}$ and $\mathcal{S}_{\rm IRT}$, the dominant entanglement entropy $\mathcal{S}_{\rm BRT}$ will transit directly to $\mathcal{S}_{\rm IRT}$.

\subsection{Page Curve} \label{Page Curve}

Now we are ready to plot the Page curve from the phase transitions among the three RT surfaces obtained in section \ref{Phase Transitions}. The Page curves for different temperatures are plotted in Fig.\ref{Page_Curve}.

At low temperature $T<T_c$, e.g.  $z_h = 0.5$ or $T=0.637$, the Page curve is plotted in Fig.\ref{Page_Curve}(a). At early times, the entanglement entropy is dominated by $\mathcal{S}_{\rm BRT}$ which increases from zero at $t=t_b$. After the first phase transition between $\mathcal{S}_{\rm BRT}$ and $\mathcal{S}_{\rm HM}$ takes place at $t-t_b\simeq 0.444$, $\mathcal{S}_{\rm HM}$ becomes dominant until the second phase transition between $\mathcal{S}_{\rm HM}$ and $\mathcal{S}_{\rm IRT}$ takes place at $t-t_b\simeq 0.759$ when $\mathcal{S}_{\rm IRT}$ becomes dominant. In this case, the Page time is determined by the phase transition between $\mathcal{S}_{\rm HM}$ and $\mathcal{S}_{\rm IRT}$ labeled by the violet dot.

As the temperature grows, the duration of $\mathcal{S}_{\rm HM}$ shrinks. At the critical temperature $T_c=2.349$ or $z_h = 0.1355$, the Page curve is plotted in Fig.\ref{Page_Curve}(b). The contribution from the HM surface completely vanishes, and $\mathcal{S}_{\rm BRT}$ transits to $\mathcal{S}_{\rm IRT}$ directly. The Page time now is determined by the phase transition between $\mathcal{S}_{\rm BRT}$ and $\mathcal{S}_{\rm IRT}$ labeled by the orange dot.

As we go beyond the critical temperature as shown in Fig.\ref{Page_Curve}(c) for the case of $T=3.183$ or $z_h = 0.1$, $\mathcal{S}_{\rm BRT}$ always transits to $\mathcal{S}_{\rm IRT}$ directly. The other two phase transitions are irrelevant. The Page time is determined by the phase transition between $\mathcal{S}_{\rm BRT}$ and $\mathcal{S}_{\rm IRT}$ labeled by the orange dot.

We conclude that the Page time decreases as the temperature grows. Among the three RT surfaces, only the IRT surface intersects with the Planck brane, and its entanglement wedge includes the interior of the eternal black hole.
According to the entanglement wedge reconstruction, we can reconstruct the interior of the black hole when $\mathcal{S}_{\rm IRT}$ is dominant. The decrease of the Page time as the temperature grows implies that one can reconstruct the interior of the black hole earlier for a higher temperature black hole. This is consistent with the intuitive expectations since higher temperature black holes evaporate faster.

\subsection{Phase Diagram} \label{Phase Diagram}

\begin{figure*}[t]
  \centering
  \subfloat[]{\includegraphics[width=0.48\textwidth]{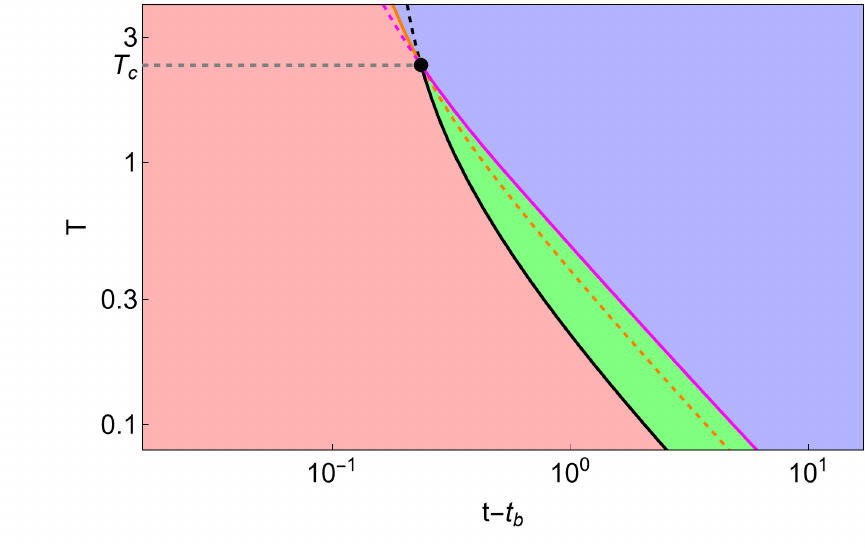}}
  \hfill
  \subfloat[]{\includegraphics[width=0.48\textwidth]{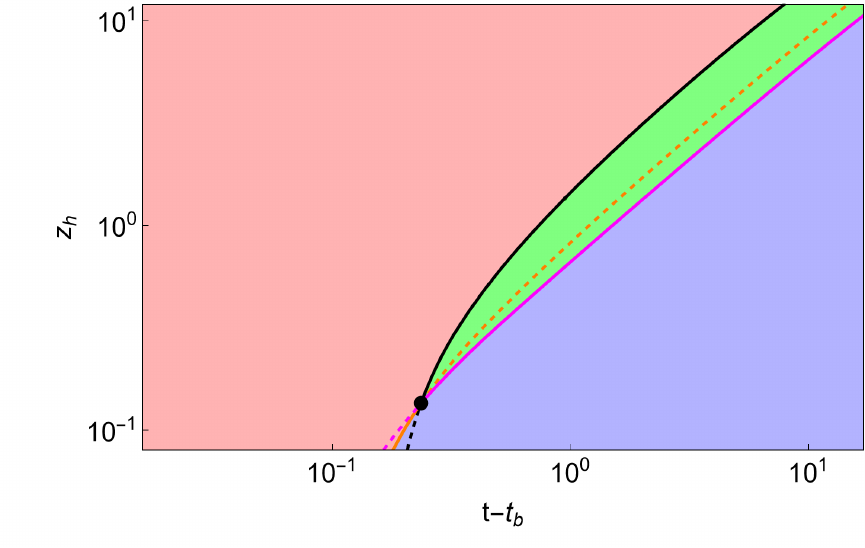}}
  \caption{Phase diagram of the Hawking radiation entanglement entropy. The red/green/blue region represents the region where $\mathcal{S}_{\rm BRT}$/$\mathcal{S}_{\rm HM}$/$\mathcal{S}_{\rm IRT}$ dominates. The black, orange, and violet lines represent the phase transitions of $\mathcal{S}_{\rm BRT}\sim \mathcal{S}_{\rm HM}$, $\mathcal{S}_{\rm BRT}\sim \mathcal{S}_{\rm IRT}$, and $\mathcal{S}_{\rm HM}\sim \mathcal{S}_{\rm IRT}$. The dashed parts do not occur in the real physical process. The black dot labels the triple point where the three phases meet together. (a) Phase diagram in the temperature $T$ and time $t-t_b$ plane. (b) Phase diagram in the black hole horizon $z_h$ and time $t-t_b$ plane.}
  \label{Phase_Diagram_b01_T&zh}
\end{figure*}

The phase diagram in temperature vs. time plane is plotted in Fig.\ref{Phase_Diagram_b01_T&zh}(a). The red region in the lower left part is dominated by $\mathcal{S}_{\rm BRT}$, the blue region in the upper right part is dominated by $\mathcal{S}_{\rm IRT}$, and the green region in between is dominated by $\mathcal{S}_{\rm HM}$. The black, orange, and violent lines are the phase boundaries between different phases. The solid parts represent the phase boundaries in the real physical process. There is a triple point at $T_c=2.349$ and $t-t_b=0.236$ (in the case of $b=0.1$) labeled by a black dot, where the three phases meet together. Below the critical temperature $T_c$, the system undergoes three phases, $\mathcal{S}_{\rm BRT}$, $\mathcal{S}_{\rm HM}$ and $\mathcal{S}_{\rm IRT}$ as time evolves, with the violet line indicating the Page time. While above the critical temperature, the system undergoes only two phases, $\mathcal{S}_{\rm BRT}$ and $\mathcal{S}_{\rm IRT}$, with the orange line indicating the Page time.

This behavior of the phase diagram is consistent with our intuitive expectation: as the temperature grows, it becomes "easier" to decode the information inside the black hole due to the earlier Page time since the black hole radiates faster at higher temperatures.

So far we have only considered the system with a fixed cutoff $b=0.1$. In the following, we will show how the phase diagram is affected by the variation of the cutoff. By using the simple relation in eq.(\ref{Temperature}), the phase diagram can be plotted in the black hole horizon vs. time plane in Fig.\ref{Phase_Diagram_b01_T&zh}(b), which will be used to illustrate the effect from varying the cutoff $b$.

\begin{figure*}[t]
	\centering
	\includegraphics[scale=0.9]{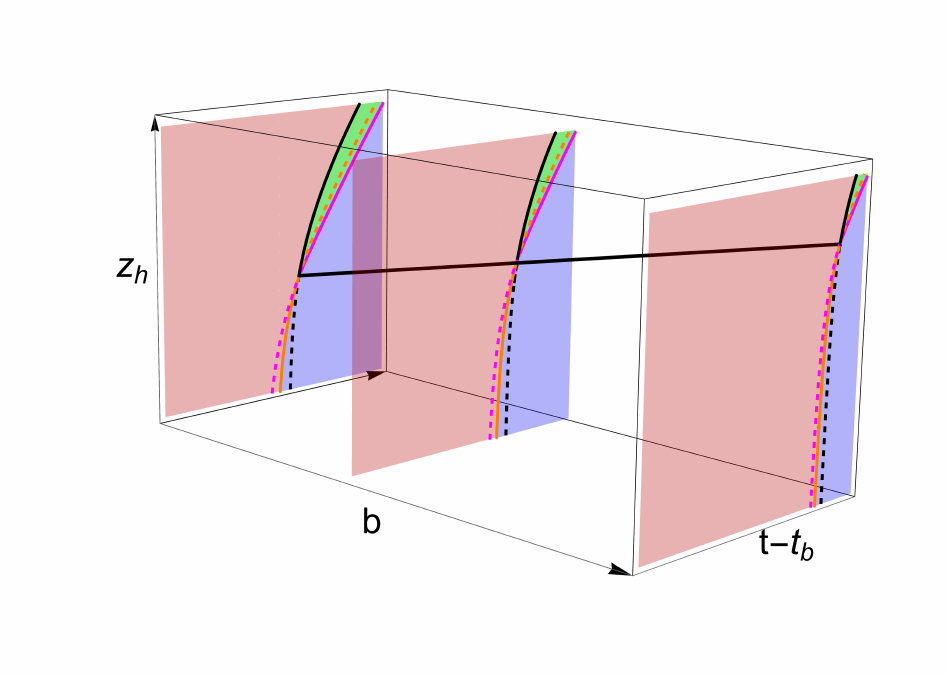}
	\caption{Different slices of the phase diagram as the cutoff $b$ is varied.}
	\label{Phase_Diagram_b}
\end{figure*}

The phase diagrams in the black hole horizon vs. time plane for three different cutoffs are plotted in Fig.\ref{Phase_Diagram_b}. We can see that, as the cutoff $b$ increases, the triple point moves to a new location with a larger black hole horizon (or lower temperature) and a later Page time. Remarkably, numerical calculations show that the path of the triple point is a straight line shown as the black line in Fig.\ref{Phase_Diagram_b}. However, the physical interpretation of this linear behavior of the triple point is not obvious to us and we will investigate it in the future.

\section{Conclusion} \label{Conclusion}

In this work, we studied the entanglement entropy in a $(d+1)$-dimensional two-sided eternal black hole system. We calculated the generalized entanglement entropy eq.(\ref{Fine-Grained Entropy}) by utilizing the doubly holographic correspondence. Our concrete setup is given by a holographic BCFT as discussed in section \ref{Double Holography}.

In our setup, we introduced two branes embedded as the boundaries of the bulk spacetime in BCFT. One is the Planck brane $\mathcal{Q}_P$ described by $x=0$, which represents the gravitational region of a radiating black hole in the doubly holographic setup. The other is the time-dependent ETW brane $\mathcal{Q}_E$ described by $x=ct$, which is the hypersurface of the earliest Hawking radiation and defines a time-dependent effective radiation region $[b_L,x_L] \cup [b_R, x_R]$.

There are three RT surfaces associated with the holographic entanglement entropy. One is the HM surface which penetrates the horizon and connects the cutoffs $b_L$ and $b_R$ directly. The HM surface increases with time due to the stretching of the interior of the black hole. The second one is the BRT surface which intersects the ETW brane $\mathcal{Q}_E$. The BRT surface increases with time due to the time-dependence of the ETW brane. The third one is the IRT surface which intersects the Planck brane $\mathcal{Q}_P$ and supports an island. The IRT surface is time-independent.

We investigated the phase transitions among the three RT surfaces in section \ref{Phase Transitions}. The phase transition time for each pair of the RT surfaces was determined by comparing their corresponding entanglement entropies. Putting all the phase transitions together, we obtained the phase diagram of the entanglement entropy as shown in Fig.\ref{Phase_Diagram_b01_T&zh}.

We found a critical temperature $T_c$ for the Page curves at different temperatures. When the temperature is lower than the critical temperature, i.e. $T<T_c$, there are three durations for the entanglement entropy as shown in Fig.\ref{Page_Curve}(a). $\mathcal{S}_{\rm BRT}$ starts from zero at the initial time $t=t_b$ and increases until the first phase transition, when $\mathcal{S}_{\rm HM}$ takes over and continuously increases. After the second phase transition, $\mathcal{S}_{\rm IRT}$ becomes dominant and remains as a constant. On the other hand, when the temperature is higher than the critical temperature, i.e. $T\ge T_c$, the duration of $\mathcal{S}_{\rm HM}$ disappears, and $\mathcal{S}_{\rm BRT}$ transits to $\mathcal{S}_{\rm IRT}$ directly as shown in Fig.\ref{Page_Curve}(b,c).

In summary, our main achievement in this work is to introduce an ETW brane $\mathcal{Q}_E$ which defines a time-dependent finite effective radiation region. The ETW brane supports a new type of RT surface, the BRT surface as discussed in section \ref{Hartman-Maldacena Surface}, which has not been considered in the previous literature. With the BRT surface, there exists a critical temperature $T_c$ as shown in the phase diagram Fig.\ref{Phase_Diagram_b01_T&zh}. At low temperature $T<T_c$, the Page time is determined by the $\mathcal{S}_{\rm HM}\sim\mathcal{S}_{\rm IRT}$ transition; while at the high temperature $T\ge T_c$, the Page time is determined by the $\mathcal{S}_{\rm BRT}\sim\mathcal{S}_{\rm IRT}$ transition. Therefore, for a high temperature eternal black hole, the true Page time is much later than that without considering the ETW brane. In this work, we only consider the eternal black hole at  fixed temperatures. In the real black hole evaporation, the temperature changes with time. It is thus important to consider this temperature-dependent phase transition to determine the Page curve. We leave this temperature-dependent effect in the real black hole evaporation for the future work.

\subsection*{Acknowledgements}
This work is supported by the Ministry of Science and Technology (MOST 109-2112-M-009-005) and National Center for Theoretical Science, Taiwan.

\bibliographystyle{unsrt}
\bibliography{Page_Curve}

\end{document}